\documentclass[aps,prd,twocolumn,superscriptaddress,nofootinbib]{revtex4-1}

\usepackage{latexsym}
\usepackage{amsmath}
\usepackage{amssymb}
\usepackage{amsfonts}
\usepackage{bm}

\usepackage{color}

\usepackage{supertabular} 
\usepackage{placeins}
\usepackage{epsfig}
\usepackage{graphicx}

\begin{document}


\title{The $\mathbf{sQ\bar{q}\bar{q}}$ $\mathbf{(q=u,\,d;\, Q=c,\,b)}$ tetraquarks in the chiral quark model}


\author{Gang Yang }
\email[]{yanggang@zjnu.edu.cn}
\affiliation{Department of Physics, Zhejiang Normal University, Jinhua 321004, China}

\author{Jialun Ping}
\email[]{jlping@njnu.edu.cn}
\affiliation{Department of Physics and Jiangsu Key Laboratory for Numerical Simulation of Large Scale Complex Systems, Nanjing Normal University, Nanjing 210023, P. R. China}

\author{Jorge Segovia}
\email[]{jsegovia@upo.es}
\affiliation{Departamento de Sistemas F\'isicos, Qu\'imicos y Naturales, \\ Universidad Pablo de Olavide, E-41013 Sevilla, Spain}



\begin{abstract}
The low-lying $sQ\bar{q}\bar{q}$ $(q=u,\,d,\, Q=c,\,b)$ tetraquark states with $J^P=0^+$, $1^+$ and $2^+$, and in the isoscalar and isovector sectors, are systematically investigated in the framework of real- and complex-scaling range of a chiral quark model, whose parameters have been fixed in advance describing hadron, hadron-hadron and multiquark phenomenology, and thus all results presented here are pure predictions. Each tetraquark configuration, compatible with the quantum numbers studied, is taken into account; this includes meson-meson, diquark-antidiquark and K-type arrangements of quarks with all possible color wave functions in four-body sector. Among the different numerical techniques to solve the Schr\"odinger-like 4-body bound state equation, we use a variational method in which the trial wave function is expanded in complex-range Gaussian basis functions, because its simplicity and flexibility. Several compact bound states and narrow resonances are found in both charm-strange $cs\bar{q}\bar{q}$ and bottom-strange $bs\bar{q}\bar{q}$ tetraquark sectors, most of them as a product of the strong coupling between the different channels. The so-called $X_{0,1}(2900)$ signals, recently found by the LHCb collaboration, are unstable in our formalism with several candidates in the single-channel computations.
\end{abstract}

\pacs{
12.38.-t \and 
12.39.-x      
}
\keywords{
Quantum Chromodynamics \and
Quark models
}

\maketitle


\section{Introduction}

The LHCb collaboration, using proton-proton collision data taken at $\sqrt{s}=7$, $8$, and $13$ TeV, with an integrated luminosity of $9\,\text{fb}^{-1}$, has recently performed an amplitude analysis~\cite{lhcb:x29001} and a model-independent study of structure~\cite{lhcb:x29002} in $B^+\rightarrow D^+D^-K^+$ decays. In both research works, in order to obtain good agreement with the data, it is found to be necessary to include new spin-$0$ and spin-$1$ charm-strange resonances in the $D^-K^+$ channel with Breit-Wigner parameters~\cite{lhcb:x29001}:
\begin{align}\nonumber
X_0(2900): \hspace*{0.2cm} M &= (2866\pm 7) \,\text{MeV} \,, \\[1ex] \nonumber
\Gamma &= (57 \pm 13) \,\text{MeV} \,, \\[1ex] \nonumber
X_1(2900): \hspace*{0.2cm} M &= (2904\pm 5) \,\text{MeV} \,, \\[1ex] \nonumber
\Gamma &= (110 \pm 12) \,\text{MeV} \,. 
\end{align}
With an overwhelming significance, these signals constitute the first clear observation of exotic hadrons that do not contain a heavy quark-antiquark pair; even more, they might be the first experimental detection of tetraquark candidates with four different flavors of quarks: $ud\bar{s}\bar{c}$. It is worth mentioning that alternative models incorporating additional hadronic effects such as re-scattering may also be able to accommodate these $D^-K^+$ structures (see, for instate, the nice review~\cite{Guo:2019twa}). More detailed experimental investigations will require larger data samples and studies of additional decay modes.

The announcement of the LHCb collaboration triggers many theoretical works using a wide variety of approaches. Before going through a brief review of them, it is fair to highlight that a couple of theoretical works indicating the possible existence of $ud\bar{s}\bar{c}$ structures were available in the literature before the release of the LHCb findings. At least the lowest-lying state, at a mass of $2.85\,\text{GeV}$, was identified using a color-magnetic interaction model~\cite{JChengPRD101140172020} and a coupled-channels unitary approach~\cite{RMolinaPRD820140102010}.

Concerning the $X_0(2900)$, many theoretical studies assign to this state total spin and parity quantum numbers $J^P=0^+$. Within this assignment, an $S$-wave $\bar{D}^*K^*(D^*\bar{K}^*)$ molecular interpretation is shared by QCD sum rules~\cite{HChen2008.07516, ssagaev200813027}, one-boson-exchange models~\cite{Mliu200807389, jhe200807782}, a quark delocalization color screening model~\cite{yxue200809516}, an effective field theory approach~\cite{RMolina200811171}, and by analyzing strong decay processes~\cite{yhuang200807959}. However, some other theoretical studies explore the possibility of having a more compact tetraquark configuration for the $X_0(2900)$ state with either di-meson or diquark-antidiquark configuration. For instance, Ref.~\cite{XHe2008.07145} uses two-body chromomagnetic interactions to find that $X_0(2900)$ can be interpreted as a radial excited $0^+$ $ud\bar{s}\bar{c}$ tetraquark. Other examples are QCD sum rules~\cite{jzhang200807295, rma200813463, zwang200807833}, and nonrelativistic and relativized quark models~\cite{gwang201009395, Garcilazo:2020bgc, mk200805993}. All mentioned theoretical works have also delivered interpretations to the $X_1(2900)$. For instance, Ref.~\cite{XHe2008.07145} concludes that the $X_1(2900)$ state is an orbitally excited compact tetraquark with $J^P=1^-$, whereas a molecular interpretation within a chiral quark model is delivered in Ref.~\cite{ytan201004045}. In any case, most theoretical works support the $J^P=1^-$ assignment~\cite{xh200807145, hc200807516, jhe200807782}.

Note, however, that debates on the nature of the $X_0(2900)$ and $X_1(2900)$ exotic states are still intense, with other alternative interpretations not yet been discarded. In an extended relativized quark model~\cite{ql200807340}, the $X_0(2900)$ signal can not be identified as a compact $ud\bar{s}\bar{c}$ tetraquark state with $J^P=0^+$. By solving the Lippmann-Schwinger equation~\cite{mh200806894}, $X_0(2900)$ can be interpreted as a $\bar{D}^*K^*$ hadronic molecule state but with quantum numbers $I(J^P)=0(1^+)$. Within an effective field theory approach~\cite{xd200911619}, the $X_1(2900)$ can not be explained as a bound state of $\bar{D}_1 K$, whereas a $D K_1$ bound state is possible at around $3.1$ GeV. Moreover, these exotic states can also be induced by triangle singularities~\cite{xliu200807190, tjb200812838}.

Additionally, decay and production properties have been also studied theoretically. Within the assumption that $X_0(2900)$ is a $S$-wave $\bar{D}^* K^*$ molecular state, its decay properties have been analyzed by using an effective field theory approach in Ref.~\cite{cxiao200914538}. The production of $X_{0,1}(2900)$ via weak decays of B-meson~\cite{ychen200901182, tburns200905352} and the hadronic effects on $X_{0,1}(2900)$ in heavy-ion collisions~\cite{labreu201014955} have been also released.

The goal of the present manuscript is to study the charm-strange tetraquark candidates seen by the LHCb collaboration using a QCD-inspired chiral quark model which has already been successfully applied in the description of other multiquark systems; \emph{e.g.} hidden-charm pentaquarks $P_c^+(4312)$, $P_c^+(4380)$, $P_c^+(4440)$ and $P_c(4457)^+$~\cite{Yang:2015bmv}, doubly-charm pentaquarks~\cite{gy:2020dcp} and doubly-heavy tetraquarks, $QQ\bar{q}\bar{q}$ $(q=u,\,d,\,s;\,Q=c,\,b)$~\cite{gy:2020dht,gy:2020dhts}. Moreover, the formulation in real- and complex-scaling method of the theoretical formalism has been discussed in detail in Ref.~\cite{YangSym2020}. The complex-scaling method (CSM) allows us to distinguish three kinds of poles: bound, resonance and scattering; and thus perform a complete analysis of the scattering singularities within the same formalism. Furthermore, the meson-meson, diquark-antidiquark and K-type configurations, plus their couplings, shall be considered for the $sc\bar{q}\bar{q}$ tetraquark system. Finally, as a natural extension, the bottom-strange tetraquarks are also studied herein.

The manuscript is arranged as follows. In Sec.~\ref{sec:model} the theoretical framework is presented; we briefly present the complex-range method applied to a chiral quark model and the $sQ\bar{q}\bar{q}$ $(q=u,\,d;\, Q=c,\,b)$ tetraquark wave-functions. Section~\ref{sec:results} is devoted to the analysis and discussion of the obtained low-lying $sQ\bar{q}\bar{q}$ $(q=u,\,d,\, Q=c,\,b)$ tetraquark states with $J^P=0^+$, $1^+$ and $2^+$, and isospin $I=0$ and $1$. Finally, we summarize and give some prospects in Sec.~\ref{sec:summary}.


\section{Theoretical framework}
\label{sec:model}

A throughout review of the theoretical formalism used herein has been recently published in Ref.~\cite{YangSym2020}. We shall, however, focused on the most relevant features of the chiral quark model and the numerical method concerning the charm(bottom)-strange tetraquarks, \emph{viz.} the $sQ\bar{q}\bar{q}$ system with $Q$ denoting either $c$- or $b$-quark, the strange quark is writte as $s$-quark and $q$ denotes the light $u$ and $d$ quarks, isospin symmetry is assumed.
 
Within the so-called complex-range studies, the relative coordinate of a two-body interaction is rotated in the complex plane by an angle $\theta$, \emph{i.e.}, $\vec{r}_{ij}\to \vec{r}_{ij} e^{i\theta}$. Therefore, the general form of the four-body Hamiltonian reads~\cite{Yang:2019itm, Yang:2020fou}:
\begin{equation}
H(\theta) = \sum_{i=1}^{4}\left( m_i+\frac{\vec{p\,}^2_i}{2m_i}\right) - T_{\text{CM}} + \sum_{j>i=1}^{4} V(\vec{r}_{ij} e^{i\theta}) \,,
\label{eq:Hamiltonian}
\end{equation}
where $m_{i}$ is the quark mass, $\vec{p}_i$ is the quark's momentum, and $T_{\text{CM}}$ is the center-of-mass kinetic energy. According to the so-called ABC theorem~\cite{JA22269, EB22280}, the complex scaled Schr\"odinger equation:
\begin{equation}\label{CSMSE}
\left[ H(\theta)-E(\theta) \right] \Psi_{JM}(\theta)=0	
\end{equation}
has (complex) eigenvalues which can be classified into three types, namely bound, resonance and scattering states. In particular, bound-states and resonances are independent of the rotated angle $\theta$, with the first ones always fixed on the coordinate-axis (there is no imaginary part of the eigenvalue), and the second ones located above the continuum threshold with a total decay width $\Gamma=-2\,\text{Im}(E)$.

The dynamics of the $sQ\bar q\bar q$ tetraquark system is driven by a two-body potential
\begin{equation}
\label{CQMV}
V(\vec{r}_{ij}) = V_{\chi}(\vec{r}_{ij}) + V_{\text{CON}}(\vec{r}_{ij}) + V_{\text{OGE}}(\vec{r}_{ij})  \,,
\end{equation}
which takes into account the most relevant features of QCD at its low energy regime: dynamical chiral symmetry breaking, confinement and the perturbative one-gluon exchange interaction. Herein, the low-lying $S$-wave positive parity $sQ\bar{q}\bar{q}$ tetraquark states shall be investigated, and thus the central and spin-spin terms of the potential are the only ones needed.

One consequence of the dynamical breaking of chiral symmetry is that Goldstone boson exchange interactions appear between constituent light quarks $u$, $d$ and $s$. Therefore, the chiral interaction can be written as~\cite{Vijande:2004he}:
\begin{equation}
V_{\chi}(\vec{r}_{ij}) = V_{\pi}(\vec{r}_{ij})+ V_{\sigma}(\vec{r}_{ij}) + V_{K}(\vec{r}_{ij}) + V_{\eta}(\vec{r}_{ij}) \,,
\end{equation}
given by
\begin{align}
&
V_{\pi}\left( \vec{r}_{ij} \right) = \frac{g_{ch}^{2}}{4\pi}
\frac{m_{\pi}^2}{12m_{i}m_{j}} \frac{\Lambda_{\pi}^{2}}{\Lambda_{\pi}^{2}-m_{\pi}
^{2}}m_{\pi} \Bigg[ Y(m_{\pi}r_{ij}) \nonumber \\
&
\hspace*{1.20cm} - \frac{\Lambda_{\pi}^{3}}{m_{\pi}^{3}}
Y(\Lambda_{\pi}r_{ij}) \bigg] (\vec{\sigma}_{i}\cdot\vec{\sigma}_{j})\sum_{a=1}^{3}(\lambda_{i}^{a}
\cdot\lambda_{j}^{a}) \,, \\
& 
V_{\sigma}\left( \vec{r}_{ij} \right) = - \frac{g_{ch}^{2}}{4\pi}
\frac{\Lambda_{\sigma}^{2}}{\Lambda_{\sigma}^{2}-m_{\sigma}^{2}}m_{\sigma} \Bigg[Y(m_{\sigma}r_{ij}) \nonumber \\
&
\hspace*{1.20cm} - \frac{\Lambda_{\sigma}}{m_{\sigma}}Y(\Lambda_{\sigma}r_{ij})
\Bigg] \,,
\end{align}
\begin{align}
& 
V_{K}\left( \vec{r}_{ij} \right)= \frac{g_{ch}^{2}}{4\pi}
\frac{m_{K}^2}{12m_{i}m_{j}}\frac{\Lambda_{K}^{2}}{\Lambda_{K}^{2}-m_{K}^{2}}m_{
K} \Bigg[ Y(m_{K}r_{ij}) \nonumber \\
&
\hspace*{1.20cm} -\frac{\Lambda_{K}^{3}}{m_{K}^{3}}Y(\Lambda_{K}r_{ij}) \Bigg] (\vec{\sigma}_{i}\cdot\vec{\sigma}_{j}) \sum_{a=4}^{7}(\lambda_{i}^{a} \cdot \lambda_{j}^{a}) \,, \\
& 
V_{\eta}\left( \vec{r}_{ij} \right) = \frac{g_{ch}^{2}}{4\pi}
\frac{m_{\eta}^2}{12m_{i}m_{j}} \frac{\Lambda_{\eta}^{2}}{\Lambda_{\eta}^{2}-m_{
\eta}^{2}}m_{\eta} \Bigg[ Y(m_{\eta}r_{ij}) \nonumber \\
&
\hspace*{1.20cm} -\frac{\Lambda_{\eta}^{3}}{m_{\eta}^{3}
}Y(\Lambda_{\eta}r_{ij}) \Bigg] (\vec{\sigma}_{i}\cdot\vec{\sigma}_{j})
\Big[\cos\theta_{p} \left(\lambda_{i}^{8}\cdot\lambda_{j}^{8}
\right) \nonumber \\
&
\hspace*{1.20cm} -\sin\theta_{p} \Big] \,,
\end{align}
where $Y(x)=e^{-x}/x$ is the standard Yukawa function. The physical $\eta$ meson, instead of the octet one, is considered by introducing the angle $\theta_p$. The $\lambda^{a}$ are the SU(3) flavor Gell-Mann matrices. Taken from their experimental values, $m_{\pi}$, $m_{K}$ and $m_{\eta}$ are the masses of the SU(3) Goldstone bosons. The value of $m_{\sigma}$ is determined through the PCAC relation $m_{\sigma}^{2}\simeq m_{\pi}^{2}+4m_{u,d}^{2}$~\cite{Scadron:1982eg}. Finally, the chiral coupling constant, $g_{ch}$, is determined from the $\pi NN$ coupling constant through
\begin{equation}
\frac{g_{ch}^{2}}{4\pi}=\frac{9}{25}\frac{g_{\pi NN}^{2}}{4\pi} \frac{m_{u,d}^{2}}{m_{N}^2} \,,
\end{equation}
which assumes that flavor SU(3) is an exact symmetry only broken by the different mass of the strange quark.

Color confinement should be encoded in the non-Abelian character of QCD. It has been demonstrated by lattice-regularized QCD that multi-gluon exchanges produce an attractive linearly rising potential proportional to the distance between infinite-heavy quarks~\cite{Bali:2005fu}. However, the spontaneous creation of light-quark pairs from the QCD vacuum may give rise at the same scale to a breakup of the created color flux-tube~\cite{Bali:2005fu}. These two observations can be described phenomenologically by
\begin{equation}
V_{\text{CON}}(\vec{r}_{ij})=\left[-a_{c}(1-e^{-\mu_{c}r_{ij}})+\Delta \right] 
(\lambda_{i}^{c}\cdot \lambda_{j}^{c}) \,,
\label{eq:conf}
\end{equation}
where $a_{c}$, $\mu_{c}$ and $\Delta$ are model parameters,\footnote{It is widely believed that confinement is flavor independent and thus it should be constraint by the light hadron spectra despite our aim is to determine energy states in heavier quark sectors~\cite{Segovia:2008zza, Segovia:2008zz}} and the SU(3) color Gell-Mann matrices are denoted as $\lambda^c$. One can see in Eq.~\eqref{eq:conf} that the potential is linear at short inter-quark distances with an effective confinement strength $\sigma = -a_{c} \, \mu_{c} \, (\lambda^{c}_{i}\cdot \lambda^{c}_{j})$, while it becomes constant at large distances, $V_{\text{thr.}} = (\Delta-a_c)(\lambda^{c}_{i}\cdot \lambda^{c}_{j})$.

Beyond the chiral symmetry breaking scale one expects the dynamics to be
governed by QCD perturbative effects. In particular, the one-gluon exchange potential, which includes the so-called coulomb and color-magnetism interactions, is the leading order contribution:
\begin{align}
&
V_{\text{OGE}}(\vec{r}_{ij}) = \frac{1}{4} \alpha_{s} (\lambda_{i}^{c}\cdot \lambda_{j}^{c}) \Bigg[\frac{1}{r_{ij}} \nonumber \\ 
&
\hspace*{1.60cm} - \frac{1}{6m_{i}m_{j}} (\vec{\sigma}_{i}\cdot\vec{\sigma}_{j}) 
\frac{e^{-r_{ij}/r_{0}(\mu_{ij})}}{r_{ij} r_{0}^{2}(\mu_{ij})} \Bigg] \,,
\end{align}
where $r_{0}(\mu_{ij})=\hat{r}_{0}/\mu_{ij}$ is a regulator which depends on the reduced mass of the $q\bar{q}$ pair, the Pauli matrices are denoted by $\vec{\sigma}$, and the contact term has been regularized as
\begin{equation}
\delta(\vec{r}_{ij}) \sim \frac{1}{4\pi r_{0}^{2}(\mu_{ij})}\frac{e^{-r_{ij} / r_{0}(\mu_{ij})}}{r_{ij}} \,.
\end{equation}

An effective scale-dependent strong coupling constant, $\alpha_s(\mu_{ij})$, provides a consistent description of mesons and baryons from light to heavy quark sectors. We use the frozen coupling constant of, for instance, Ref.~\cite{Segovia:2013wma}
\begin{equation}
\alpha_{s}(\mu_{ij})=\frac{\alpha_{0}}{\ln\left(\frac{\mu_{ij}^{2}+\mu_{0}^{2}}{\Lambda_{0}^{2}} \right)} \,,
\end{equation}
in which $\alpha_{0}$, $\mu_{0}$ and $\Lambda_{0}$ are parameters of the model.

The model parameters are listed in Table~\ref{tab:model}. They have been fixed in advance reproducing hadron~\cite{Segovia:2009zz, Segovia:2011zza, Segovia:2015dia, Yang:2017xpp, Yang:2019lsg}, hadron-hadron ~\cite{Ortega:2016mms, Ortega:2016pgg, Ortega:2017qmg, Ortega:2018cnm, Ortega:2020uvc} and multiquark~\cite{Vijande:2006jf, Yang:2015bmv, Yang:2017rpg, Yang:2018oqd, Yang:2020twg} phenomenology. Additionally, for later concern, Table~\ref{MesonMass} lists theoretical and experimental (if available) masses of $1S$ and $2S$ states of $K^{(*)}$, $D^{(*)}$ and $B^{(*)}$ mesons.

\begin{table}[!t]
\caption{\label{tab:model} Model parameters.}
\begin{ruledtabular}
\begin{tabular}{llr}
Quark masses     & $m_q\,(q=u,\,d)$ (MeV) & 313 \\
                 & $m_s$ (MeV) &  555 \\
                 & $m_c$ (MeV) & 1752 \\
                 & $m_b$ (MeV) & 5100 \\[2ex]
Goldstone bosons & $\Lambda_\pi=\Lambda_\sigma~$ (fm$^{-1}$) &   4.20 \\
                 & $\Lambda_\eta=\Lambda_K$ (fm$^{-1}$)      &   5.20 \\
                 & $g^2_{ch}/(4\pi)$                         &   0.54 \\
                 & $\theta_P(^\circ)$                        & -15 \\[2ex]
Confinement      & $a_c$ (MeV)         & 430 \\
                 & $\mu_c$ (fm$^{-1})$ & 0.70 \\
                 & $\Delta$ (MeV)      & 181.10 \\[2ex]
OGE              & $\alpha_0$              & 2.118 \\
                 & $\Lambda_0~$(fm$^{-1}$) & 0.113 \\
                 & $\mu_0~$(MeV)           & 36.976 \\
                 & $\hat{r}_0~$(MeV~fm)    & 28.170 \\
\end{tabular}
\end{ruledtabular}
\end{table}

\begin{table}[!t]
\caption{\label{MesonMass} Theoretical and experimental (if available) masses of $nL=1S$ and $2S$ states of $K^{(*)}$, $D^{(*)}$ and $B^{(*)}$ mesons.}
\begin{ruledtabular}
\begin{tabular}{lccc}
Meson & $nL$ & $M_{\text{The.}}$ (MeV) & $M_{\text{Exp.}}$ (MeV) \\
\hline
$K$ & $1S$ &  $481$ & $494$ \\
    & $2S$ & $1468$ & - \\[2ex]
$K^*$ & $1S$ &  $907$ & $892$ \\
      & $2S$ & $1621$ & - \\[2ex]
$D$ & $1S$ & $1897$ & $1870$ \\
    & $2S$ & $2648$ & - \\[2ex]
$D^*$ & $1S$ & $2017$ & $2007$ \\
      & $2S$ & $2704$ & - \\[2ex]
$B$ & $1S$ & $5278$ & $5280$ \\
    & $2S$ & $5984$ & - \\[2ex]
$B^*$ & $1S$ & $5319$ & $5325$ \\
      & $2S$ & $6005$ & - 
\end{tabular}
\end{ruledtabular}
\end{table}

\begin{figure}[ht]
\epsfxsize=3.4in \epsfbox{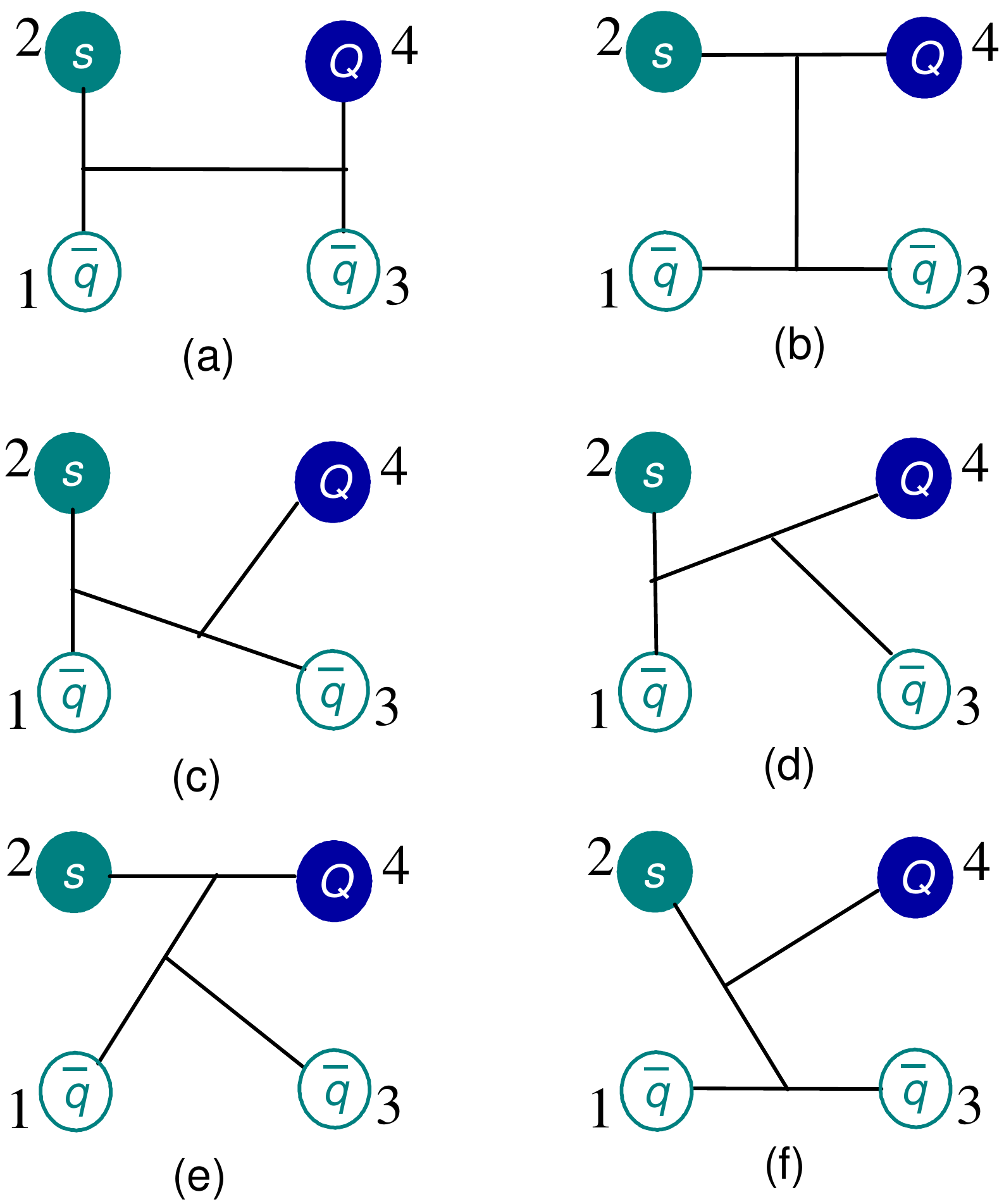}
\caption{Six types of configurations in $sQ\bar{q}\bar{q}$ $(q=u,\,d,\,Q=c,\,b)$ tetraquarks. Panel $(a)$ is meson-meson structure, panel $(b)$ is diquark-antidiquark one and the other K-type structures are from panel $(c)$ to $(f)$.} \label{QQqq}
\end{figure}

Figure~\ref{QQqq} shows six kinds of configurations for the $sQ\bar{q}\bar{q}$ tetraquark system. In particular, Fig.~\ref{QQqq}(a) is the meson-meson structure, Fig.~\ref{QQqq}(b) is the diquark-antidiquark one, and the other K-type configurations are from panels (c) to (f). All of them, and their couplings, are considered in our investigation. However, for the purpose of solving a manageable 4-body problem, the K-type configurations are sometimes restricted. It is important to note herein that just one configuration would be enough for the calculation, if all radial and orbital excited states were taken into account; however, this is obviously much less efficient and thus an economic way is to combine the different configurations in the ground state to perform the calculation.

The multiquark system's wave function at the quark level is an internal product of color, spin, flavor and space terms. Concerning the color degree-of-freedom, the colorless wave function of a 4-quark system in meson-meson configuration, as illustrated in Fig.~\ref{QQqq}(a), can be obtained by either two coupled color-singlet clusters, $1\otimes 1$:
\begin{align}
\label{Color1}
\chi^c_1 &= \frac{1}{3}(\bar{r}r+\bar{g}g+\bar{b}b)\times (\bar{r}r+\bar{g}g+\bar{b}b) \,,
\end{align}
or two coupled color-octet clusters, $8\otimes 8$:
\begin{align}
\label{Color2}
\chi^c_2 &= \frac{\sqrt{2}}{12}(3\bar{b}r\bar{r}b+3\bar{g}r\bar{r}g+3\bar{b}g\bar{g}b+3\bar{g}b\bar{b}g+3\bar{r}g\bar{g}r
\nonumber\\
&+3\bar{r}b\bar{b}r+2\bar{r}r\bar{r}r+2\bar{g}g\bar{g}g+2\bar{b}b\bar{b}b-\bar{r}r\bar{g}g
\nonumber\\
&-\bar{g}g\bar{r}r-\bar{b}b\bar{g}g-\bar{b}b\bar{r}r-\bar{g}g\bar{b}b-\bar{r}r\bar{b}b) \,.
\end{align}
The first color state is the so-called color-singlet channel and the second one is the named hidden-color case.

The color wave functions associated to the diquark-antidiquark structure shown in Fig.~\ref{QQqq}(b) are the coupled color triplet-antitriplet clusters, $3\otimes \bar{3}$:
\begin{align}
\label{Color3}
\chi^c_3 &= \frac{\sqrt{3}}{6}(\bar{r}r\bar{g}g-\bar{g}r\bar{r}g+\bar{g}g\bar{r}r-\bar{r}g\bar{g}r+\bar{r}r\bar{b}b
\nonumber\\
&-\bar{b}r\bar{r}b+\bar{b}b\bar{r}r-\bar{r}b\bar{b}r+\bar{g}g\bar{b}b-\bar{b}g\bar{g}b
\nonumber\\
&+\bar{b}b\bar{g}g-\bar{g}b\bar{b}g) \,,
\end{align}
and the coupled color sextet-antisextet clusters, $6\otimes \bar{6}$:
\begin{align}
\label{Color4}
\chi^c_4 &= \frac{\sqrt{6}}{12}(2\bar{r}r\bar{r}r+2\bar{g}g\bar{g}g+2\bar{b}b\bar{b}b+\bar{r}r\bar{g}g+\bar{g}r\bar{r}g
\nonumber\\
&+\bar{g}g\bar{r}r+\bar{r}g\bar{g}r+\bar{r}r\bar{b}b+\bar{b}r\bar{r}b+\bar{b}b\bar{r}r
\nonumber\\
&+\bar{r}b\bar{b}r+\bar{g}g\bar{b}b+\bar{b}g\bar{g}b+\bar{b}b\bar{g}g+\bar{g}b\bar{b}g) \,.
\end{align}

Meanwhile, the colorless wave functions of the K-type structures shown in Fig.~\ref{QQqq}(c) to (f) are 
\begin{align}
\label{Color5}
\chi^c_5 &= \frac{1}{6\sqrt{2}}(\bar{r}r\bar{r}r+\bar{g}g\bar{g}g-2\bar{b}b\bar{b}b)+
\nonumber\\
&\frac{1}{2\sqrt{2}}(\bar{r}b\bar{b}r+\bar{r}g\bar{g}r+\bar{g}b\bar{b}g+\bar{g}r\bar{r}g+\bar{b}g\bar{g}b+\bar{b}r\bar{r}b)-
\nonumber\\
&\frac{1}{3\sqrt{2}}(\bar{g}g\bar{r}r+\bar{r}r\bar{g}g)+\frac{1}{6\sqrt{2}}(\bar{b}b\bar{r}r+\bar{b}b\bar{g}g+\bar{r}r\bar{b}b+\bar{g}g\bar{b}b) \,,
\end{align}
\begin{align}
\label{Color6}
\chi^c_6 &= \chi^c_1 \,,
\end{align}
\begin{align}
\label{Color7}
\chi^c_7 &= \chi^c_1 \,,
\end{align}
\begin{align}
\label{Color8}
\chi^c_8 &= \frac{1}{4}(1-\frac{1}{\sqrt{6}})\bar{r}r\bar{g}g-\frac{1}{4}(1+\frac{1}{\sqrt{6}})\bar{g}g\bar{g}g-\frac{1}{4\sqrt{3}}\bar{r}g\bar{g}r+
\nonumber\\
&\frac{1}{2\sqrt{2}}(\bar{r}b\bar{b}r+\bar{g}b\bar{b}g+\bar{b}g\bar{g}b+\bar{g}r\bar{r}g+\bar{b}r\bar{r}b)+
\nonumber\\
&\frac{1}{2\sqrt{6}}(\bar{r}r\bar{b}b-\bar{g}g\bar{b}b+\bar{b}b\bar{g}g+\bar{g}g\bar{r}r-\bar{b}b\bar{r}r) \,,
\end{align}
\begin{align}
\label{Color9}
\chi^c_9 &= \frac{1}{2\sqrt{6}}(\bar{r}b\bar{b}r+\bar{r}r\bar{b}b+\bar{g}b\bar{b}g+\bar{g}g\bar{b}b+\bar{r}g\bar{g}r+\bar{r}r\bar{g}g+
\nonumber\\
&\bar{b}b\bar{g}g+\bar{b}g\bar{g}b+\bar{g}g\bar{r}r+\bar{g}r\bar{r}g+\bar{b}b\bar{r}r+\bar{b}r\bar{r}b)+
\nonumber\\
&\frac{1}{\sqrt{6}}(\bar{r}r\bar{r}r+\bar{g}g\bar{g}g+\bar{b}b\bar{b}b) \,,
\end{align}
\begin{align}
\label{Color10}
\chi^c_{10} &= \frac{1}{2\sqrt{3}}(\bar{r}b\bar{b}r-\bar{r}r\bar{b}b+\bar{g}b\bar{b}g-\bar{g}g\bar{b}b+\bar{r}g\bar{g}r-\bar{r}r\bar{g}g-
\nonumber\\
&\bar{b}b\bar{g}g+\bar{b}g\bar{g}b-\bar{g}g\bar{r}r+\bar{g}r\bar{r}g-\bar{b}b\bar{r}r+\bar{b}r\bar{r}b) \,,
\end{align}
\begin{align}
\label{Color11}
\chi^c_{11} &= \chi^c_9 \,,
\end{align}
\begin{align}
\label{Color12}
\chi^c_{12} &= -\chi^c_{10} \,.
\end{align}

As for the flavor degree-of-freedom, since the quark content of the investigated 4-quark system is $sQ\bar{q}\bar{q}$, both isoscalar $I=0$ and isovector $I=1$ sectors will be discussed. Note also that if the flavor wave-function is denoted as $\chi^{fi}_{I, M_I}$, the superscript $i=1$ and $2$ will refer to $sc\bar{q}\bar{q}$ and $sb\bar{q}\bar{q}$ systems, respectively. The specific wave functions read as below
\begin{align}
&
\chi_{0,0}^{fi} = \frac{1}{\sqrt{2}}(\bar{u}s\bar{d}Q-\bar{d}s\bar{u}Q) \,, \\
&
\chi_{1,0}^{fi} = \frac{1}{\sqrt{2}}(\bar{u}s\bar{d}Q+\bar{d}s\bar{u}Q) \,,
\end{align}
where the third component of the isospin, $M_I$, is fixed to be zero for simplicity since the Hamiltonian does not have a flavour-dependent interaction which can distinguish the third component of the isospin quantum number.

We are going to considered $S$-wave ground states with spin ranging from from $S=0$ to $2$. Therefore, the spin wave functions, $\chi^{\sigma_i}_{S, M_S}$, are given by ($M_S$ can be set to be equal to $S$ without loss of generality):
\begin{align}
\label{SWF1}
\chi_{0,0}^{\sigma_{u_1}}(4) &= \chi^\sigma_{00}\chi^\sigma_{00} \,, \\
\chi_{0,0}^{\sigma_{u_2}}(4) &= \frac{1}{\sqrt{3}}(\chi^\sigma_{11}\chi^\sigma_{1,-1}-\chi^\sigma_{10}\chi^\sigma_{10}+\chi^\sigma_{1,-1}\chi^\sigma_{11}) \,, \\
\chi_{0,0}^{\sigma_{u_3}}(4) &= \frac{1}{\sqrt{2}}\big((\sqrt{\frac{2}{3}}\chi^\sigma_{11}\chi^\sigma_{\frac{1}{2}, -\frac{1}{2}}-\sqrt{\frac{1}{3}}\chi^\sigma_{10}\chi^\sigma_{\frac{1}{2}, \frac{1}{2}})\chi^\sigma_{\frac{1}{2}, -\frac{1}{2}} \nonumber \\ 
&-(\sqrt{\frac{1}{3}}\chi^\sigma_{10}\chi^\sigma_{\frac{1}{2}, -\frac{1}{2}}-\sqrt{\frac{2}{3}}\chi^\sigma_{1, -1}\chi^\sigma_{\frac{1}{2}, \frac{1}{2}})\chi^\sigma_{\frac{1}{2}, \frac{1}{2}}\big) \,, \\
\chi_{0,0}^{\sigma_{u_4}}(4) &= \frac{1}{\sqrt{2}}(\chi^\sigma_{00}\chi^\sigma_{\frac{1}{2}, \frac{1}{2}}\chi^\sigma_{\frac{1}{2}, -\frac{1}{2}}-\chi^\sigma_{00}\chi^\sigma_{\frac{1}{2}, -\frac{1}{2}}\chi^\sigma_{\frac{1}{2}, \frac{1}{2}}) \,,
\end{align}
\begin{align}
\chi_{1,1}^{\sigma_{w_1}}(4) &= \chi^\sigma_{00}\chi^\sigma_{11} \,, \\ 
\chi_{1,1}^{\sigma_{w_2}}(4) &= \chi^\sigma_{11}\chi^\sigma_{00} \,, \\
\chi_{1,1}^{\sigma_{w_3}}(4) &= \frac{1}{\sqrt{2}} (\chi^\sigma_{11} \chi^\sigma_{10}-\chi^\sigma_{10} \chi^\sigma_{11}) \,, \\
\chi_{1,1}^{\sigma_{w_4}}(4) &= \sqrt{\frac{3}{4}}\chi^\sigma_{11}\chi^\sigma_{\frac{1}{2}, \frac{1}{2}}\chi^\sigma_{\frac{1}{2}, -\frac{1}{2}}-\sqrt{\frac{1}{12}}\chi^\sigma_{11}\chi^\sigma_{\frac{1}{2}, -\frac{1}{2}}\chi^\sigma_{\frac{1}{2}, \frac{1}{2}} \nonumber \\ 
&-\sqrt{\frac{1}{6}}\chi^\sigma_{10}\chi^\sigma_{\frac{1}{2}, \frac{1}{2}}\chi^\sigma_{\frac{1}{2}, \frac{1}{2}} \,, \\
\chi_{1,1}^{\sigma_{w_5}}(4) &= (\sqrt{\frac{2}{3}}\chi^\sigma_{11}\chi^\sigma_{\frac{1}{2}, -\frac{1}{2}}-\sqrt{\frac{1}{3}}\chi^\sigma_{10}\chi^\sigma_{\frac{1}{2}, \frac{1}{2}})\chi^\sigma_{\frac{1}{2}, \frac{1}{2}} \,, \\
\chi_{1,1}^{\sigma_{w_6}}(4) &= \chi^\sigma_{00}\chi^\sigma_{\frac{1}{2}, \frac{1}{2}}\chi^\sigma_{\frac{1}{2}, \frac{1}{2}} \,, \\
\label{SWF2}
\chi_{2,2}^{\sigma_{1}}(4) &= \chi^\sigma_{11}\chi^\sigma_{11} \,.
\end{align}
The superscripts $u_1,\ldots,u_4$ and $w_1,\ldots,w_6$ determine the spin wave function for each configuration of the $sQ\bar q\bar q$ tetraquark system, their specific values are shown in Table~\ref{SpinIndex}. Furthermore, the expressions above are obtained by considering the coupling of two sub-clusters whose spin wave functions are given by trivial SU(2) algebra, and the necessary basis reads as
\begin{align}
\label{Spin}
\chi^\sigma_{11} &= \chi^\sigma_{\frac{1}{2}, \frac{1}{2}} \chi^\sigma_{\frac{1}{2}, \frac{1}{2}} \,, \\
\chi^\sigma_{1,-1} &= \chi^\sigma_{\frac{1}{2}, -\frac{1}{2}} \chi^\sigma_{\frac{1}{2}, -\frac{1}{2}} \,, \\
\chi^\sigma_{10} &= \frac{1}{\sqrt{2}}(\chi^\sigma_{\frac{1}{2}, \frac{1}{2}} \chi^\sigma_{\frac{1}{2}, -\frac{1}{2}}+\chi^\sigma_{\frac{1}{2}, -\frac{1}{2}} \chi^\sigma_{\frac{1}{2}, \frac{1}{2}}) \,, \\
\chi^\sigma_{00} &= \frac{1}{\sqrt{2}}(\chi^\sigma_{\frac{1}{2}, \frac{1}{2}} \chi^\sigma_{\frac{1}{2}, -\frac{1}{2}}-\chi^\sigma_{\frac{1}{2}, -\frac{1}{2}} \chi^\sigma_{\frac{1}{2}, \frac{1}{2}}) \,, 
\end{align}

\begin{table}[!t]
\caption{\label{SpinIndex} The values of the superscripts $u_1,\ldots,u_4$ and $w_1,\ldots,w_6$ that determine the spin wave function for each configuration of the $sQ\bar q\bar q$ tetraquark system.}
\begin{ruledtabular}
\begin{tabular}{lcccccc}
& Di-meson & Diquark-antidiquark & $K_1$ & $K_2$ & $K_3$ & $K_4$ \\
\hline
$u_1$ & 1 & 3 & & & & \\
$u_2$ & 2 & 4 & & & & \\
$u_3$ &   &   & 5 & 7 &  9 & 11 \\
$u_4$ &   &   & 6 & 8 & 10 & 12 \\[2ex]
$w_1$ & 1 & 4 & & & & \\
$w_2$ & 2 & 5 & & & & \\
$w_3$ & 3 & 6 & & & & \\
$w_4$ &   &   & 7 & 10 & 13 & 16 \\
$w_5$ &   &   & 8 & 11 & 14 & 17 \\
$w_6$ &   &   & 9 & 12 & 15 & 18
\end{tabular}
\end{ruledtabular}
\end{table}

Among the different methods to solve the Schr\"odinger-like 4-body bound state equation, we use the Rayleigh-Ritz variational principle which is one of the most extended tools to solve eigenvalue problems because its simplicity and flexibility. Moreover, we use the complex-range method and thus the spatial wave function is written as follows:
\begin{equation}
\label{eq:WFexp}
\psi_{LM_L}(\theta)= \left[ \left[ \phi_{n_1l_1}(\vec{\rho}e^{i\theta}\,) \phi_{n_2l_2}(\vec{\lambda}e^{i\theta}\,)\right]_{l} \phi_{n_3l_3}(\vec{R}e^{i\theta}\,) \right]_{L M_L} \,,
\end{equation}
where the internal Jacobi coordinates are defined as
\begin{align}
\vec{\rho} &= \vec{x}_1-\vec{x}_2 \,, \\
\vec{\lambda} &= \vec{x}_3 - \vec{x}_4 \,, \\
\vec{R} &= \frac{m_1 \vec{x}_1 + m_2 \vec{x}_2}{m_1+m_2}- \frac{m_3 \vec{x}_3 + m_4 \vec{x}_4}{m_3+m_4} \,,
\end{align}
for the meson-meson configuration of Fig.~\ref{QQqq}(a); and as
\begin{align}
\vec{\rho} &= \vec{x}_1-\vec{x}_3 \,, \\
\vec{\lambda} &= \vec{x}_2 - \vec{x}_4 \,, \\
\vec{R} &= \frac{m_1 \vec{x}_1 + m_3 \vec{x}_3}{m_1+m_3}- \frac{m_2 \vec{x}_2 + m_4 \vec{x}_4}{m_2+m_4} \,,
\end{align}
for the diquark-antdiquark structure of Fig.~\ref{QQqq}(b). The remaining K-type configurations shown in Fig.~\ref{QQqq}(c) to \ref{QQqq}(f) are ($i, j, k, l$ take values according to the panels (c) to (f) of Fig.~\ref{QQqq}):
\begin{align}
\vec{\rho} &= \vec{x}_i-\vec{x}_j \,, \\
\vec{\lambda} &= \vec{x}_k- \frac{m_i \vec{x}_i + m_j \vec{x}_j}{m_i+m_j} \,, \\
\vec{R} &= \vec{x}_l- \frac{m_i \vec{x}_i + m_j \vec{x}_j+m_k \vec{x}_k}{m_i+m_j+m_k} \,.
\end{align}
It becomes obvious now that the center-of-mass kinetic term $T_{CM}$ can be completely eliminated for a non-relativistic system defined in any of the above sets of relative coordinates.

A crucial aspect of the Rayleigh-Ritz variational method is the basis expansion of the trial wave function. We are going to use the Gaussian expansion method (GEM)~\cite{Hiyama:2003cu} in which each relative coordinate is expanded in terms of Gaussian basis functions whose sizes are taken in geometric progression. This method has proven to be very efficient on solving the bound-state problem of a multiquark systems~\cite{Yang:2015bmv, Yang:2018oqd, gy:2020dcp, gy:2020dht} and the details on how the geometric progression is fixed can be found in \emph{e.g} Ref.~\cite{Yang:2015bmv}. Therefore, the form of the orbital wave functions, $\phi$'s, in Eq.~\eqref{eq:WFexp} is 
\begin{align}
&
\phi_{nlm}(\vec{r}e^{i\theta}\,) = N_{nl} (re^{i\theta})^{l} e^{-\nu_{n} (re^{i\theta})^2} Y_{lm}(\hat{r}) \,.
\end{align}
Since only $S$-wave states of charm(bottom)-strange tetraquarks are investigated in this work, no laborious Racah algebra is needed while computing matrix elements. In this case, the value of the spherical harmonic function is just a constant, \emph{viz.} $Y_{00}=\sqrt{1/4\pi}$.

Finally, the complete wave-function that fulfills the Pauli principle is written as
\begin{equation}
\label{TPs}
\Psi_{JM_J,I,i,j,k}(\theta)={\cal A} \left[ \left[ \psi_{L}(\theta) \chi^{\sigma_i}_{S}(4) \right]_{JM_J} \chi^{f_j}_I \chi^{c}_k \right] \,,
\end{equation}
where $\cal{A}$ is the antisymmetry operator of $sQ\bar{q}\bar{q}$ tetraquarks which takes into account the fact of having two identical light anti-quarks, $(\bar{q}\bar{q})$. Its definition, according to Fig.~\ref{QQqq}, is 
\begin{equation}
\label{Antisym}
{\cal{A}} = 1-(13) \,.
\end{equation}
This is necessary since the complete wave function of the 4-quark system is constructed from two sub-clusters: meson-meson, diquark-antidiquark and K-type configurations.


\section{Results}
\label{sec:results}

In the present calculation, we investigate all possible $S$-wave charm(bottom)-strange tetraquarks by taking into account di-meson, diquark-antidiquark and K-type configurations. In our approach, a $sQ\bar{q}\bar{q}$ tetraquark state has positive parity assuming that the angular momenta $l_1$, $l_2$ and $l_3$ in Eq.~\eqref{eq:WFexp} are all equal to zero. Accordingly, the total angular momentum, $J$, coincides with the total spin, $S$, and can take values of $0$, $1$ and $2$. Besides, the value of isospin, $I$, can be either $0$ or $1$ considering the quark content of the $sQ\bar{q}\bar{q}$ system.

Tables~\ref{GresultCC1} to~\ref{GresultCC12} list our calculated results of the lowest-lying $sc\bar{q}\bar{q}$ and $sb\bar{q}\bar{q}$ tetraquark states. The allowed meson-meson, diquark-antidiquark and K-type configurations are listed in the first column; when possible, the experimental value of the non-interacting meson-meson threshold is labeled in parentheses. Each channel is assigned an index in the 2nd column, it reflects a particular combination of spin ($\chi_J^{\sigma_i}$), flavor ($\chi_I^{f_j}$) and color ($\chi_k^c$) wave functions that are shown explicitly in the 3rd column. The theoretical mass obatined in each channel is shown in the 4th column and the coupled result for each kind of configuration is presented in the 5th column. When a complete coupled-channels calculation is performed, last row of the table indicates the lowest-lying mass and binding energy; moreover, their dominant components and the distances between any two quarks of the system are also discussed. When the CSM is used in the complete coupled-channels calculation, we show in Figs.~\ref{PP1} to~\ref{PP12}, the distribution of complex eigen-energies and, therein, the obtained bound and resonance states are indicated inside circles.

Let us proceed now to describe in detail our theoretical findings for each sector of $sQ\bar{q}\bar{q}$ tetraquarks, with $Q$ either a $c$- or a $b$-quark.

\subsection{The $\mathbf{sc\bar{q}\bar{q}}$ tetraquarks}

There are two bound states with quantum numbers $I(J^P)=0(0^+)$ and $0(1^+)$, respectively. Moreover, several narrow resonances whose masses cluster around 3.1 GeV and 3.5 GeV are found in the remaining $I(J^P)$ channels studied herein. Each isoscalar and vector sectors with total spin and parity $J^P=0^+$, $1^+$ and $2^+$ shall be discussed individually below.

\begin{table}[!t]
\caption{\label{GresultCC1} Lowest-lying $sc\bar{q}\bar{q}$ tetraquark states with $I(J^P)=0(0^+)$ calculated within the real range formulation of the chiral quark model.
The allowed meson-meson, diquark-antidiquark and K-type configurations are listed in the first column; when possible, the experimental value of the non-interacting meson-meson threshold is labeled in parentheses. Each channel is assigned an index in the 2nd column, it reflects a particular combination of spin ($\chi_J^{\sigma_i}$), flavor ($\chi_I^{f_j}$) and color ($\chi_k^c$) wave functions that are shown explicitly in the 3rd column. The theoretical mass obatined in each channel is shown in the 4th column and the coupled result for each kind of configuration is presented in the 5th column.
When a complete coupled-channels calculation is performed, last row of the table indicates the lowest-lying mass and binding energy.
(unit: MeV).}
\begin{ruledtabular}
\begin{tabular}{lcccc}
~~Channel   & Index & $\chi_J^{\sigma_i}$;~$\chi_I^{f_j}$;~$\chi_k^c$ & $M$ & Mixed~~ \\
        &   &$[i; ~j; ~k]$ &  \\[2ex]
$(D K)^1 (2364)$          & 1  & [1;~1;~3]  & $2378$ & \\
$(D^* K^*)^1 (2899)$  & 2  & [2;~1;~3]   & $2924$ & $2374$ \\[2ex]
$(D K)^8$      & 3  & [1;~1;~4]   & $3198$ & \\
$(D^* K^*)^8$  & 4  & [2;~1;~4]   & $3014$ & $2861$ \\[2ex]
$(sc)(\bar{q}\bar{q})$      & 5   & [3;~1;~1]  & $2735$ & \\
$(sc)^*(\bar{q}\bar{q})^*$  & 6  & [4;~1;~2]   & $2997$ & $2632$ \\[2ex]
$K_1$  & 7  & [5;~1;~5]   & $2899$ & \\
  & 8  & [5;~1;~6]   & $3048$ & \\
  & 9  & [6;~1;~5]   & $3163$ & \\
  & 10  & [6;~1;~6]   & $2630$ & $2561$ \\[2ex]
$K_2$  & 11  & [7;~1;~7]   & $2964$ & \\
  & 12  & [7;~1;~8]   & $3011$ & \\
  & 13  & [8;~1;~7]   & $2491$ & \\
  & 14  & [8;~1;~8]   & $3183$ & $2460$ \\[2ex]
$K_3$  & 15  & [9;~1;~9]   & $2978$ & \\
  & 16  & [9;~1;~10]   & $3755$ & \\
  & 17  & [10;~1;~9]   & $3777$ & \\
  & 18  & [10;~1;~10]   & $2871$ & $2722$ \\[2ex]
$K_4$  & 19  & [11;~1;~11]   & $2988$ & \\
  & 20  & [12;~1;~12]   & $2739$ & $2611$ \\[2ex]
\multicolumn{4}{c}{Complete coupled-channels:} & $2342$ \\
   &   &    &  & $E_B=-36$
\end{tabular}
\end{ruledtabular}
\end{table}

\begin{table}[!t]
\caption{\label{GresultComp1} Relative strengths of various meson-meson, diquark-antidiquark and K-type components in the wave function of the $I(J^P)=0(0^+)$ $sc\bar{q}\bar{q}$ tetraquark bound-state obtained when a complete coupled-channels calculation is performed. The superscripts $1$ and $8$ are for singlet-color and hidden-color channels, respectively.}
\begin{ruledtabular}
\begin{tabular}{lcccc}
 ~~$(DK)^1$~~  & ~~$(D^*K^*)^1$~~   & ~~$(DK)^8$~~ &
 ~~$(D^*K^*)^8$~~ & ~~$(sc)(\bar{q}\bar{q})$~~ \\
 ~~13.5\%~~  & ~~9.9\%~~  & ~~2.8\%  ~~& 2.6\% ~~ & 4.7\%~~\\[2ex]
 ~~$(sc)^*(\bar{q}\bar{q})^*$~~ & ~~$K_1$~~ & ~~$K_2$~~ & ~~$K_3$~~ & ~~$K_4$~~ \\ 
 ~~0.8\%~~  & ~~19.4\%~~  & ~~35.0\%  ~~& 6.7\% ~~ & 4.6\%~~\\
\end{tabular}
\end{ruledtabular}
\end{table}

\begin{table}[!t]
\caption{\label{tab:dis1} The distance, in fm, between any two quarks of the $I(J^P)=0(0^+)$ $sc\bar{q}\bar{q}$ tetraquark bound-state obtained when a complete coupled-channels calculation is performed.}
\begin{ruledtabular}
\begin{tabular}{cccc}
  ~~$r_{\bar{q}\bar{q}}$ & $r_{\bar{q}s}$ & $r_{\bar{q}c}$ & $r_{sc}$~~  \\[2ex]
  ~~1.08 & 0.84 & 0.82 & 0.98~~ \\
\end{tabular}
\end{ruledtabular}
\end{table}

\begin{figure}[ht]
\includegraphics[width=0.49\textwidth, trim={2.3cm 2.0cm 2.0cm 1.0cm}]{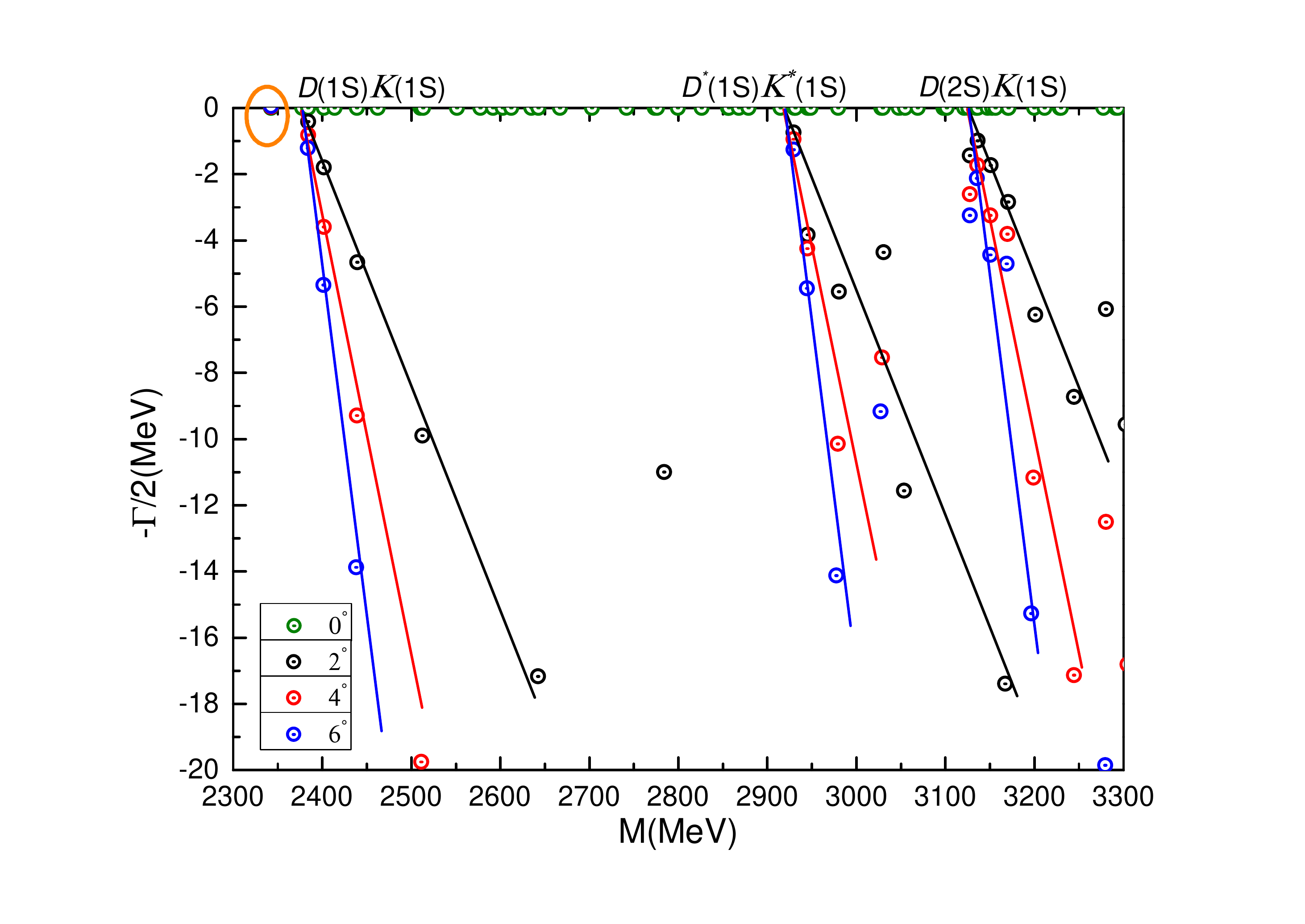}
\caption{\label{PP1} The complete coupled-channels calculation of the $sc\bar{q}\bar{q}$ tetraquark system with $I(J^P)=0(0^+)$ quantum numbers. We use the complex-scaling method of the chiral quark model varying $\theta$ from $0^\circ$ to $6^\circ$.}
\end{figure}

{\bf The $\bm{I(J^P)=0(0^+)}$ sector:} Two meson-meson channels, $DK$ and $D^* K^*$ in both color-singlet and hidden-color configurations, two diquark-antidiquark channels, $(sc)(\bar{q}\bar{q})$ and $(sc)^*(\bar{q}\bar{q})^*$, along with K-type configurations are individually studied in Table~\ref{GresultCC1}. One can see that the formation of a bound state is not possible in each single channel calculation. The two singlet-color meson-meson states are located above their respective thresholds with masses at 2.38 GeV and 2.92 GeV, respectively. The other tetraquark configurations: hidden-color meson-meson, diquark-antidiquark, and K-type channels, present states in an energy region which ranges from 2.7 GeV to 3.7 GeV. If a coupled-channels calculation among each type of configuration is performed, a loosely bound state is found in the singlet color meson-meson channel with a mass of 2374~MeV and a binding energy of $-4$~MeV. Furthermore, we observe a strong coupling effect in the complete coupled-channels calculation which helps in obtaining a tightly bound state whose mass and binding energy are 2342~MeV and $-36$~MeV, respectively.

We have investigated the composition of this state in Table~\ref{GresultComp1}. The dominant structure is of K-type contributing 55\% to the total wave function, but traces of singlet-color $DK$ (14\%) and $D^*K^*$ (10\%) components are also found. Table~\ref{tab:dis1} indicates that all quark--(anti-)quark pairs are separated about the same distance with a value around 1 fm. All of these features indicate that this state might be a compact $sc\bar q\bar q$ tetraquark.

In a further step, the complete coupled-channels calculation is performed using the complex-range method and its output is shown in Fig.~\ref{PP1}. Therein, with a rotated angle ranging from $0^\circ$ to $6^\circ$, the bound state is fixed at 2342~MeV of real-axis, and we mark it with an orange circle. One can also see in Fig.~\ref{PP1} that the scattering states of $D(1S)K(1S)$, $D^*(1S)K^*(1S)$ and $D(2S)K(1S)$ are well presented within an interval of 2.3-3.3 GeV. No resonance pole is found around 2.9 GeV within the complete coupled-channels calculation; nevertheless, several good candidates for the $X_0(2900)$ signal reported by the LHCb collaboration are found in, for instance, single channel calculations of K-type configurations: 2.89 GeV in $K_1$, 2.96 GeV in $K_2$ and 2.97 GeV in $K_3$; and the hidden-color coupled-channels calculation with a tentative mass of 2.86 GeV.


\begin{table}[!t]
\caption{\label{GresultCC2} Lowest-lying $sc\bar{q}\bar{q}$ tetraquark states with $I(J^P)=0(1^+)$ calculated within the real range formulation of the chiral quark model.
The allowed meson-meson, diquark-antidiquark and K-type configurations are listed in the first column; when possible, the experimental value of the non-interacting meson-meson threshold is labeled in parentheses. Each channel is assigned an index in the 2nd column, it reflects a particular combination of spin ($\chi_J^{\sigma_i}$), flavor ($\chi_I^{f_j}$) and color ($\chi_k^c$) wave functions that are shown explicitly in the 3rd column. The theoretical mass obatined in each channel is shown in the 4th column and the coupled result for each kind of configuration is presented in the 5th column.
When a complete coupled-channels calculation is performed, last row of the table indicates the lowest-lying mass and binding energy.
(unit: MeV).}
\scalebox{0.949}{
\begin{ruledtabular}
\begin{tabular}{lcccc}
~~Channel   & Index & $\chi_J^{\sigma_i}$;~$\chi_I^{f_j}$;~$\chi_k^c$ & $M$ & Mixed~~ \\
        &   &$[i; ~j; ~k]$ &  \\[2ex]
$(D^* K)^1 (2501)$      & 1    & [1;~1;~3]   & $2498$ & \\
$(D K^*)^1 (2762)$      & 2    & [2;~1;~3]    & $2804$ & \\
$(D^* K^*)^1 (2899)$  & 3    & [3;~1;~3]    & $2924$ & $2497$ \\[2ex]
$(D^* K)^8$          & 4    & [1;~1;~4]    & $3224$ & \\
$(D K^*)^8$          & 5    & [2;~1;~4]    & $3269$ & \\
$(D^* K^*)^8$     & 6    & [3;~1;~4]    & $3232$ & $2935$ \\[2ex]
$(sc)(\bar{q}\bar{q})^*$   & 7    & [4;~1;~1]    & $2766$ &  \\
$(sc)^*(\bar{q}\bar{q})$   & 8    & [5;~1;~2]    & $3180$ &  \\
$(sc)^*(\bar{q}\bar{q})^*$   & 9    & [6;~1;~2]    & $3068$ & $2697$ \\[2ex]
$K_1$  & 10    & [7;~1;~5]    & $3275$ & \\
  & 11    & [7;~1;~6]    & $3158$ & \\
  & 12    & [8;~1;~5]    & $2971$ & \\
  & 13    & [8;~1;~6]    & $3049$ & \\
  & 14    & [9;~1;~5]    & $3184$ & \\
  & 15    & [9;~1;~6]    & $2689$ & $2636$ \\[2ex]
$K_2$  & 16    & [10;~1;~7]    & $2932$ & \\
  & 17    & [10;~1;~8]    & $2659$ & \\
  & 18    & [11;~1;~7]    & $3016$ & \\
  & 19    & [11;~1;~8]    & $3269$ & \\
  & 20    & [12;~1;~7]    & $2571$ & \\
  & 21    & [12;~1;~8]    & $3209$ & $2554$ \\[2ex]
$K_3$  & 22    & [13;~1;~9]    & $3069$ & \\
  & 23    & [13;~1;~10]    & $2945$ & \\
  & 24    & [14;~1;~9]    & $3062$ & \\
  & 25    & [14;~1;~10]    & $2951$ & \\
  & 26    & [15;~1;~9]    & $3171$ & \\
  & 27    & [15;~1;~10]    & $3788$ & $2815$ \\[2ex]
$K_4$  & 28    & [16;~1;~11]    & $3188$ & \\
  & 29    & [17;~1;~11]    & $3037$ & \\
  & 30    & [18;~1;~12]    & $2762$ & $2668$ \\[2ex]
\multicolumn{4}{c}{Complete coupled-channels:} & $2467$\\
   &   &    &  & $E_B=-31$
\end{tabular}
\end{ruledtabular}
}
\end{table}

\begin{table}[!t]
\caption{\label{GresultComp2} Relative strengths of various meson-meson, diquark-antidiquark and K-type components in the wave function of the $I(J^P)=0(1^+)$ $sc\bar{q}\bar{q}$ tetraquark bound-state obtained when a complete coupled-channels calculation is performed. The superscripts $1$ and $8$ are for singlet-color and hidden-color channels, respectively.}
\begin{ruledtabular}
\begin{tabular}{ccccc}
  ~~$(D^*K)^1$~~  & ~~$(DK^*)^1$~~   & ~~$(D^*K^*)^1$~~ &
  ~~$(D^*K)^8$~~  & ~~$(DK^*)^8$~~  \\
 ~~2.9\%~~  & ~~0.8\%~~  & ~~0.7\%  ~~& 3.4\% ~~ & 2.6\%~~\\[2ex]
  ~~$(D^*K^*)^8$~~ & ~~$(sc)(\bar{q}\bar{q})^*$~~  & ~~$(sc)^*(\bar{q}\bar{q})$~~   & ~~$(sc)^*(\bar{q}\bar{q})^*$~~ & ~~$K_1$~~ \\
 ~~2.2\%~~  & ~~1.3\%~~  & ~~0.2\%  ~~& 0.6\% ~~ & 21.7\%~~\\[2ex]
 ~~$K_2$~~ & ~~$K_3$~~ & ~~$K_4$~~ &  &  \\ 
 ~~12.4\%~~  & ~~49.7\%~~  & ~~1.5\%  ~~&  & \\
\end{tabular}
\end{ruledtabular}
\end{table}

\begin{table}[!t]
\caption{\label{tab:dis2} The distance, in fm, between any two quarks of the $I(J^P)=0(1^+)$ $sc\bar{q}\bar{q}$ tetraquark bound-state obtained when a complete coupled-channels calculation is performed.}
\begin{ruledtabular}
\begin{tabular}{cccc}
  ~~$r_{\bar{q}\bar{q}}$ & $r_{\bar{q}s}$ & $r_{\bar{q}c}$ & $r_{sc}$~~  \\[2ex]
  ~~1.15 & 0.89 & 0.90 & 1.06~~ \\
\end{tabular}
\end{ruledtabular}
\end{table}

\begin{figure}[ht]
\includegraphics[width=0.49\textwidth, trim={2.3cm 2.0cm 2.0cm 1.0cm}]{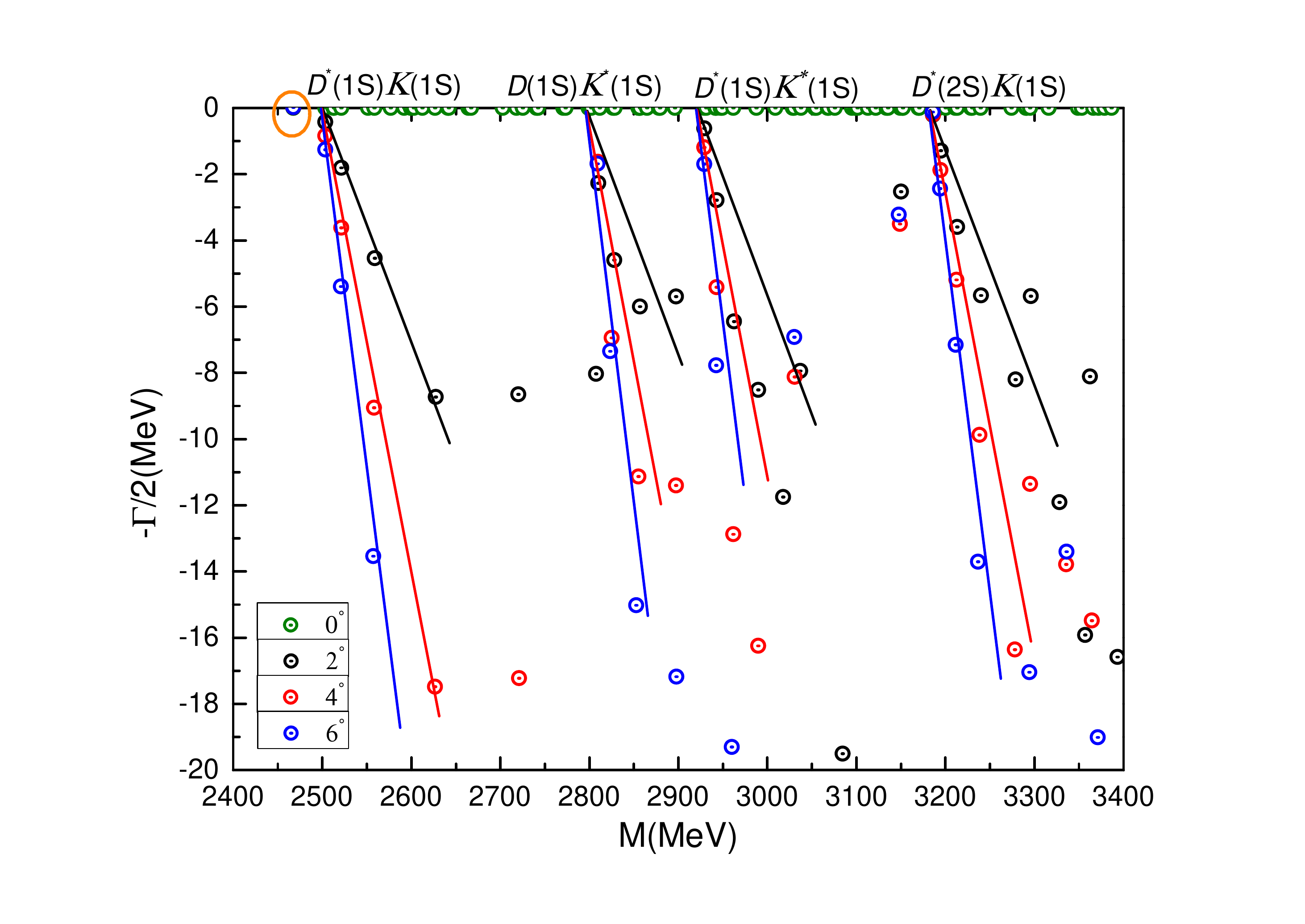}
\caption{\label{PP2} The complete coupled-channels calculation of the $sc\bar{q}\bar{q}$ tetraquark system with $I(J^P)=0(1^+)$ quantum numbers. We use the complex-scaling method of the chiral quark model varying $\theta$ from $0^\circ$ to $6^\circ$.}
\end{figure}

{\bf The $\bm{I(J^P)=0(1^+)}$ sector:} There are 30 channels in this case which include three color-singlet meson-meson configurations, another three in the hidden-color meson-meson ones, three more in diquark-antidiquark arrangement, and 21 K-type configurations. Table~\ref{GresultCC2} shows the obtained masses in each channel which are always above the lowest meson-meson threshold. The three di-meson structures in color-singlet configuration have masses at 2.50 GeV, 2.80 GeV and 2.92 GeV, respectively. Furthermore, masses of the hidden-color meson-meson configurations are $\sim$3.2 GeV; diquark-antidiquark channels are characterized with masses around 3.1 GeV, except for the $(sc)(\bar{q}\bar{q})$ channel with a mass of 2.77 GeV; and, for the K-type configurations, each calculated channel presents a mass lying in an interval of 2.5-3.3 GeV.

If one considers the coupling between channels of the same kind of tetraquark configuration, the lowest masses are located at 2497~MeV, 2935~MeV, 2697~MeV, 2636~MeV, 2554~MeV, 2815~MeV and 2668~MeV, respectively. Particularly interesting is the weakly bound state of color-singlet di-meson configuration. This result is quite similar to the one obtained in the $I(J^P)=0(0^+)$ $cs\bar{q}\bar{q}$ tetraquark sector. Meanwhile, a deeper binding energy, $E_B=-31$~MeV, is obtained when a complete coupled-channels calculation is performed, the predicted bound state has a mass M=2467~MeV.

We have investigated the composition of the last state in Table~\ref{GresultComp2}. The overwhelming dominant structure is of K-type, contributing approximately 85\% to the total wave function. Table~\ref{tab:dis2} indicates that all quark--(anti-)quark pairs are separated about the same distance with a value around 1 fm. All of these features indicate that this state might be a compact $sc\bar q\bar q$ tetraquark.

In a further step, the complete coupled-channels calculation is performed using the complex-range method and its output is shown in Fig.~\ref{PP2}. Therein, with a rotated angle ranging from $0^\circ$ to $6^\circ$, the bound state is fixed at 2467~MeV of real-axis, and we mark it with an orange circle. One can also see in Fig.~\ref{PP2} that the scattering states of $D^*(1S)K(1S)$, $D(1S)K^*(1S)$, $D^*(1S)K^*(1S)$ and $D^*(2S)K(1S)$ are well presented within an interval of 2.4-3.4 GeV. No resonance pole is found around 2.9 GeV within the complete coupled-channels calculation; nevertheless, a good candidate for the $X_1(2900)$ signal reported by the LHCb collaboration is found in the hidden-color coupled-channels calculation with a tentative mass of 2.94 GeV.


\begin{table}[!t]
\caption{\label{GresultCC3} Lowest-lying $sc\bar{q}\bar{q}$ tetraquark states with $I(J^P)=0(2^+)$ calculated within the real range formulation of the chiral quark model.
The allowed meson-meson, diquark-antidiquark and K-type configurations are listed in the first column; when possible, the experimental value of the non-interacting meson-meson threshold is labeled in parentheses. Each channel is assigned an index in the 2nd column, it reflects a particular combination of spin ($\chi_J^{\sigma_i}$), flavor ($\chi_I^{f_j}$) and color ($\chi_k^c$) wave functions that are shown explicitly in the 3rd column. The theoretical mass obatined in each channel is shown in the 4th column and the coupled result for each kind of configuration is presented in the 5th column.
When a complete coupled-channels calculation is performed, last row of the table indicates the lowest-lying mass and binding energy.
(unit: MeV).}
\begin{ruledtabular}
\begin{tabular}{lccc}
~~Channel   & Index & $\chi_J^{\sigma_i}$;~$\chi_I^{f_j}$;~$\chi_k^c$ & $M$~~ \\
        &   &$[i; ~j; ~k]$ &  \\[2ex]
$(D^* K^*)^1 (2899)$  & 1   & [1;~1;~3]  & $2924$  \\[2ex]
$(D^* K^*)^8$  & 2   & [1;~1;~3]   & $3308$   \\[2ex]
$(sc)^*(\bar{q}\bar{q})^*$  & 3   & [1;~1;~3]   & $3182$  \\[2ex]
$K_1$  & 4  & [1;~1;~3]   & $3295$  \\
            & 5   & [1;~1;~3]   & $3182$  \\[2ex]
$K_2$  & 6  & [1;~1;~3]   & $3041$  \\
            & 7   & [1;~1;~3]   & $3295$  \\[2ex]
$K_3$  & 8  & [1;~1;~3]   & $3175$  \\
            & 9   & [1;~1;~3]   & $3800$  \\[2ex]
$K_4$  & 10 & [1;~1;~3]   & $3216$  \\[2ex]
\multicolumn{3}{c}{Complete coupled-channels:} & $2924$
\end{tabular}
\end{ruledtabular}
\end{table}

\begin{figure}[ht]
\includegraphics[width=0.49\textwidth, trim={2.3cm 2.0cm 2.0cm 1.0cm}]{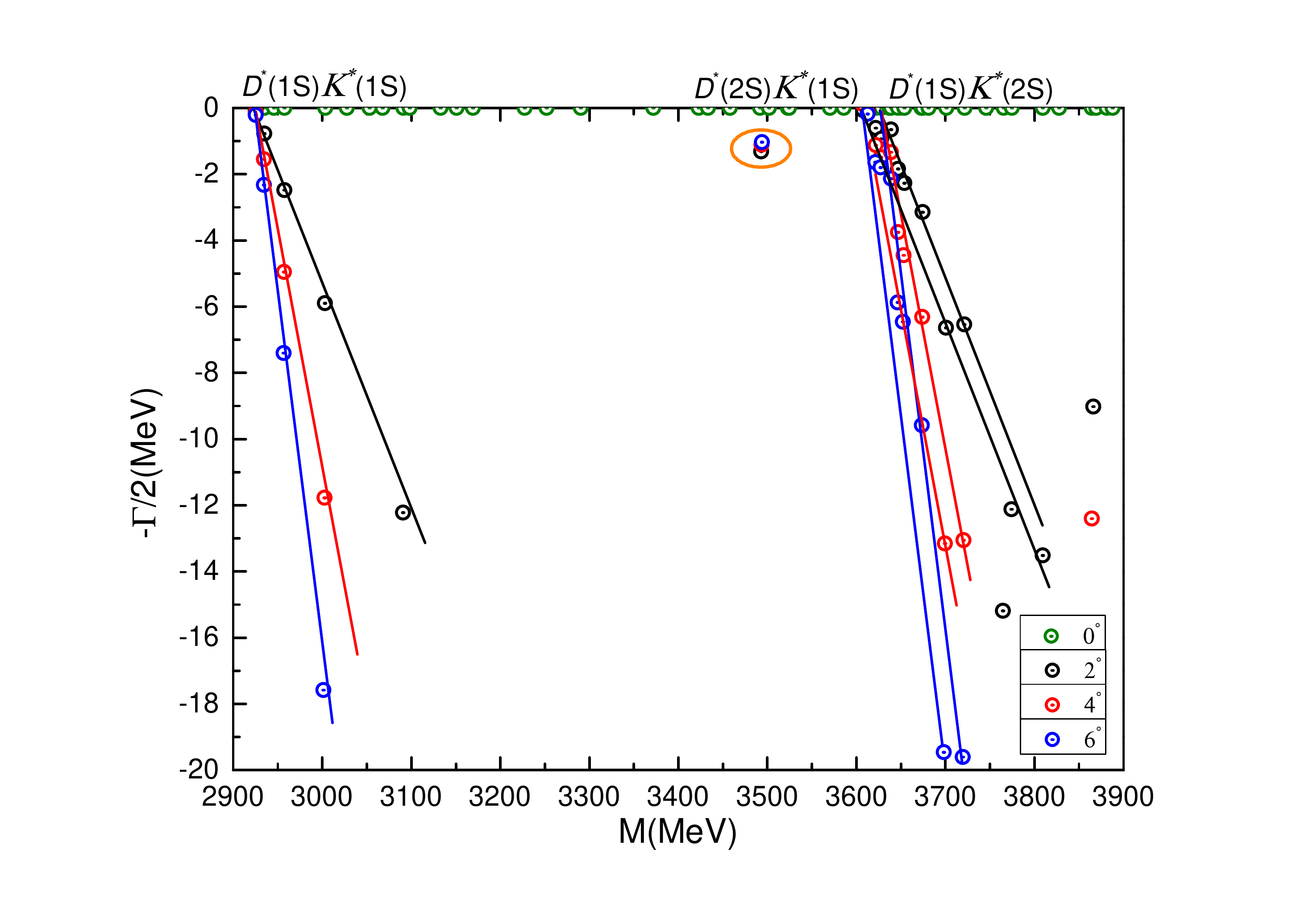}
\caption{\label{PP3} The complete coupled-channels calculation of the $sc\bar{q}\bar{q}$ tetraquark system with $I(J^P)=0(2^+)$ quantum numbers. We use the complex-scaling method of the chiral quark model varying $\theta$ from $0^\circ$ to $6^\circ$.}
\end{figure}

{\bf The $\bm{I(J^P)=0(2^+)}$ state:} Table~\ref{GresultCC3} shows that only one $D^*K^*$ meson-meson structure (in both color-singlet and hidden-color configurations), one $(sc)^*(\bar{q}\bar{q})^*$ diquark-antidiquark arrangement and seven K-type configurations contribute to the $I(J^P)=0(2^+)$ $sc\bar q\bar q$ tetraquark state. Firstly, for this highest spin state, no bound state is found neither in each single channel calculation nor in any coupled-channels ones. The mass of the color-singlet $D^*K^*$ channel is 2924~MeV, which is just at the theoretical non-interacting meson-meson threshold value. The other exotic configurations: hidden-color di-meson, diquark-antidiquark and K-type, are all located above 3.0 GeV. Even when a complete coupled-channels computation is performed, the scattering nature of states remain unchanged.

We can also notice that the nature of $I(J^P)=0(2^+)$ $sc\bar{q}\bar{q}$ tetraquark seems different with respect the $0(0^+)$ and $0(1^+)$ cases. Figure.~\ref{PP3} shows the distribution of complex energies in the complete coupled-channels calculation, the scattering nature of the $D^*(1S)K^*(1S)$, $D^*(2S)K^*(1S)$ and $D^*(1S)K^*(2S)$ states can be clearly identified. Moreover, the $D^*(2S)K^*(1S)(3611)$ and $D^*(1S)K^*(2S)(3638)$ structures are almost degenerated. In the complex plane of Fig.~\ref{PP3}, a fixed dot is found at 3493~MeV with a width of 2.6~MeV, indicating the existence of a $D^*(1S)K^*(1S)$ molecular state.


\begin{table}[!t]
\caption{\label{GresultCC4} Lowest-lying $sc\bar{q}\bar{q}$ tetraquark states with $I(J^P)=1(0^+)$ calculated within the real range formulation of the chiral quark model.
The allowed meson-meson, diquark-antidiquark and K-type configurations are listed in the first column; when possible, the experimental value of the non-interacting meson-meson threshold is labeled in parentheses. Each channel is assigned an index in the 2nd column, it reflects a particular combination of spin ($\chi_J^{\sigma_i}$), flavor ($\chi_I^{f_j}$) and color ($\chi_k^c$) wave functions that are shown explicitly in the 3rd column. The theoretical mass obatined in each channel is shown in the 4th column and the coupled result for each kind of configuration is presented in the 5th column.
When a complete coupled-channels calculation is performed, last row of the table indicates the lowest-lying mass and binding energy.
(unit: MeV).}
\begin{ruledtabular}
\begin{tabular}{lcccc}
~~Channel   & Index & $\chi_J^{\sigma_i}$;~$\chi_I^{f_j}$;~$\chi_k^c$ & $M$ & Mixed~~ \\
        &   &$[i; ~j; ~k]$ &  \\[2ex]
$(D K)^1 (2364)$          & 1  & [1;~1;~3]  & $2378$ & \\
$(D^* K^*)^1 (2899)$  & 2  & [2;~1;~3]   & $2924$ & $2378$ \\[2ex]
$(D K)^8$      & 3  & [1;~1;~4]   & $3314$ & \\
$(D^* K^*)^8$  & 4  & [2;~1;~4]   & $3294$ & $3014$ \\[2ex]
$(sc)(\bar{q}\bar{q})$      & 5   & [3;~1;~1]  & $3214$ & \\
$(sc)^*(\bar{q}\bar{q})^*$  & 6  & [4;~1;~2]   & $3086$ & $3035$ \\[2ex]
$K_1$  & 7  & [5;~1;~5]   & $3257$ & \\
  & 8  & [5;~1;~6]   & $3219$ & \\
  & 9  & [6;~1;~5]   & $3288$ & \\
  & 10  & [6;~1;~6]   & $2785$ & $2773$ \\[2ex]
$K_2$  & 11  & [7;~1;~7]   & $3076$ & \\
  & 12  & [7;~1;~8]   & $3279$ & \\
  & 13  & [8;~1;~7]   & $2617$ & \\
  & 14  & [8;~1;~8]   & $3322$ & $2604$ \\[2ex]
$K_3$  & 15  & [9;~1;~9]   & $3666$ & \\
  & 16  & [9;~1;~10]   & $3085$ & \\
  & 17  & [10;~1;~9]   & $3207$ & \\
  & 18  & [10;~1;~10]   & $3781$ & $3012$ \\[2ex]
$K_4$  & 19  & [11;~1;~11]   & $3072$ & \\
  & 20  & [12;~1;~12]   & $3253$ & $3016$ \\[2ex]
\multicolumn{4}{c}{Complete coupled-channels:} & $2378$
\end{tabular}
\end{ruledtabular}
\end{table}

\begin{figure}[ht]
\includegraphics[width=0.49\textwidth, trim={2.3cm 2.0cm 2.0cm 1.0cm}]{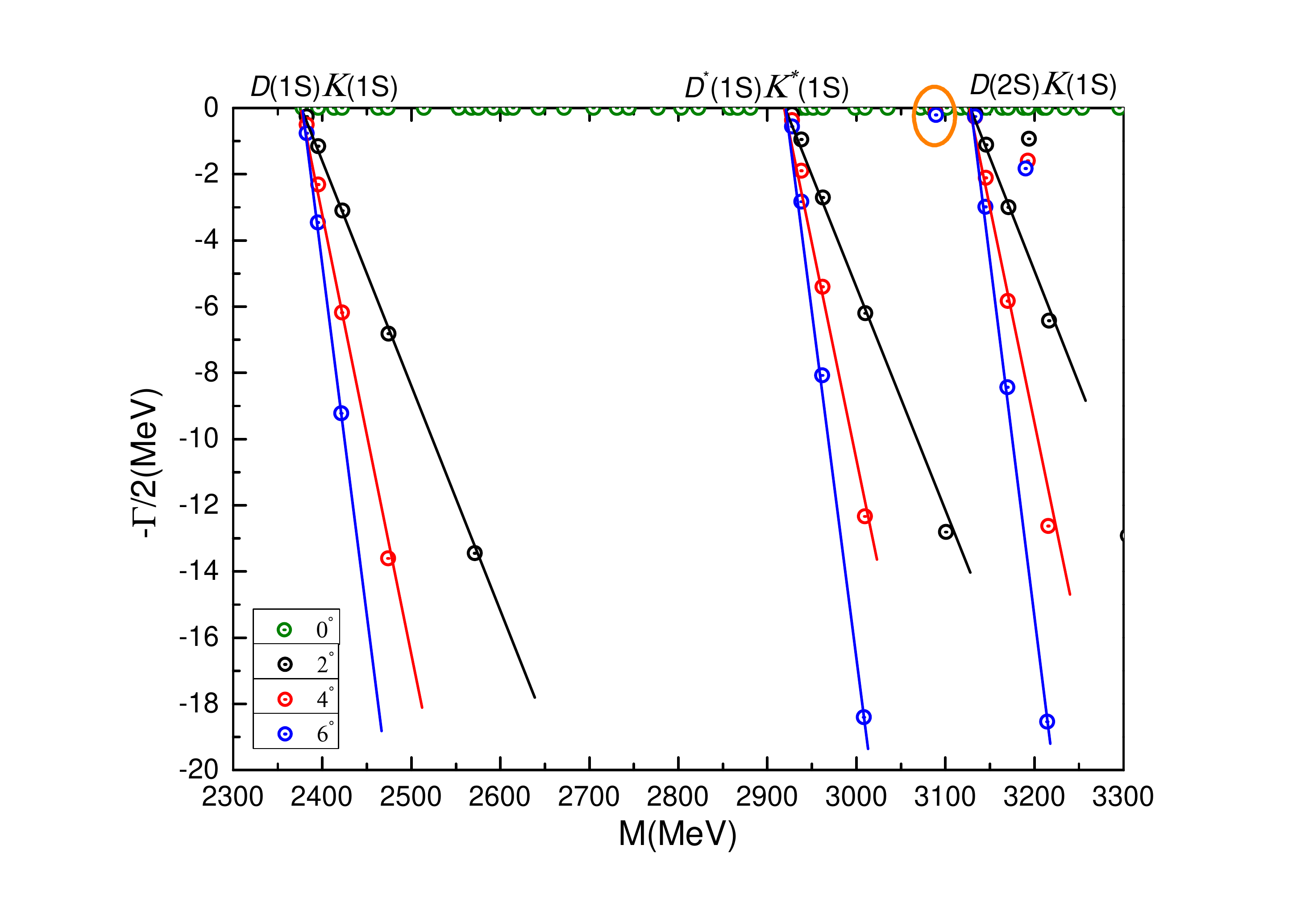}
\caption{\label{PP4} The complete coupled-channels calculation of the $sc\bar{q}\bar{q}$ tetraquark system with $I(J^P)=1(0^+)$ quantum numbers. We use the complex-scaling method of the chiral quark model varying $\theta$ from $0^\circ$ to $6^\circ$.}
\end{figure}

{\bf The $\bm{I(J^P)=1(0^+)}$ sector:} Table~\ref{GresultCC4} lists our results for the isovector $sc\bar{q}\bar{q}$ tetraquark with quantum numbers $J^P=0^+$. As in the case of the $I(J^P)=0(0^+)$ state, 20 channels are under investigation and no one shows a bound state. Masses of color-singlet meson-meson configurations are 2378~MeV and 2924~MeV, respectively. The other channels are generally in a mass region between 3.0 GeV and 3.7 GeV, except for two channels, $K_1$ and $K_2$ configurations, which present masses 2785~MeV and 2617~MeV. Additionally, this output is not changed when we perform coupled-channels calculations in either each configuration's sector and considering all channels.

The complete coupled-channels calculation has been extended to the complex-range. Although bound states are unavailable, one narrow resonance is found. Figure~\ref{PP4} shows the distribution of complex energies within the energy range 2.3-3.3 GeV. The $D(1S)K(1S)$, $D^*(1S)K^*(1S)$ and $D(2S)K(1S)$ scattering states are clearly shown. Moreover, one can find an unchanged pole which lies between $D^*(1S)K^*(1S)$ and $D(2S)K(1S)$ threshold lines, and it is extremely close to the real-axis. The calculated mass and width are 3089~MeV and 0.3~MeV, respectively. The dominant channel is $D^*(1S)K^*(1S)$ for this resonance pole and it is far from $D(1S)K(1S)$ threshold. Besides, due to the quite narrow width, this resonance state should be stable against two mesons strong decay processes. 

\begin{table}[!t]
\caption{\label{GresultCC5} Lowest-lying $sc\bar{q}\bar{q}$ tetraquark states with $I(J^P)=1(1^+)$ calculated within the real range formulation of the chiral quark model.
The allowed meson-meson, diquark-antidiquark and K-type configurations are listed in the first column; when possible, the experimental value of the non-interacting meson-meson threshold is labeled in parentheses. Each channel is assigned an index in the 2nd column, it reflects a particular combination of spin ($\chi_J^{\sigma_i}$), flavor ($\chi_I^{f_j}$) and color ($\chi_k^c$) wave functions that are shown explicitly in the 3rd column. The theoretical mass obatined in each channel is shown in the 4th column and the coupled result for each kind of configuration is presented in the 5th column.
When a complete coupled-channels calculation is performed, last row of the table indicates the lowest-lying mass and binding energy.
(unit: MeV).}
\scalebox{0.97}{\begin{ruledtabular}
\begin{tabular}{lcccc}
~~Channel   & Index & $\chi_J^{\sigma_i}$;~$\chi_I^{f_j}$;~$\chi_k^c$ & $M$ & Mixed~~ \\
        &   &$[i; ~j; ~k]$ &  \\[2ex]
$(D^* K)^1 (2501)$      & 1    & [1;~1;~3]   & $2498$ & \\
$(D K^*)^1 (2762)$      & 2    & [2;~1;~3]    & $2804$ & \\
$(D^* K^*)^1 (2899)$  & 3    & [3;~1;~3]    & $2924$ & $2498$ \\[2ex]
$(D^* K)^8$          & 4    & [1;~1;~4]    & $3288$ & \\
$(D K^*)^8$          & 5    & [2;~1;~4]    & $3271$ & \\
$(D^* K^*)^8$     & 6    & [3;~1;~4]    & $3279$ & $3043$ \\[2ex]
$(sc)(\bar{q}\bar{q})^*$   & 7    & [4;~1;~1]    & $3210$ &  \\
$(sc)^*(\bar{q}\bar{q})$   & 8    & [5;~1;~2]    & $3154$ &  \\
$(sc)^*(\bar{q}\bar{q})^*$   & 9    & [6;~1;~2]    & $3112$ & $3068$ \\[2ex]
$K_1$  & 10    & [7;~1;~5]    & $3122$ & \\
  & 11    & [7;~1;~6]    & $3124$ & \\
  & 12    & [8;~1;~5]    & $3260$ & \\
  & 13    & [8;~1;~6]    & $3200$ & \\
  & 14    & [9;~1;~5]    & $3267$ & \\
  & 15    & [9;~1;~6]    & $2806$ & $2793$ \\[2ex]
$K_2$  & 16    & [10;~1;~7]    & $3042$ & \\
  & 17    & [10;~1;~8]    & $3288$ & \\
  & 18    & [11;~1;~7]    & $3047$ & \\
  & 19    & [11;~1;~8]    & $3174$ & \\
  & 20    & [12;~1;~7]    & $2666$ & \\
  & 21    & [12;~1;~8]    & $3295$ & $2644$ \\[2ex]
$K_3$  & 22    & [13;~1;~9]    & $3191$ & \\
  & 23    & [13;~1;~10]    & $3128$ & \\
  & 24    & [14;~1;~9]    & $3183$ & \\
  & 25    & [14;~1;~10]    & $3129$ & \\
  & 26    & [15;~1;~9]    & $3779$ & \\
  & 27    & [15;~1;~10]    & $3156$ & $3051$ \\[2ex]
$K_4$  & 28    & [16;~1;~11]    & $3126$ & \\
  & 29    & [17;~1;~11]    & $3082$ & \\
  & 30    & [18;~1;~12]    & $3245$ & $3046$ \\[2ex]
\multicolumn{4}{c}{Complete coupled-channels:} & $2498$
\end{tabular}
\end{ruledtabular}}
\end{table}

\begin{figure}[ht]
\includegraphics[width=0.49\textwidth, trim={2.3cm 2.0cm 2.0cm 1.0cm}]{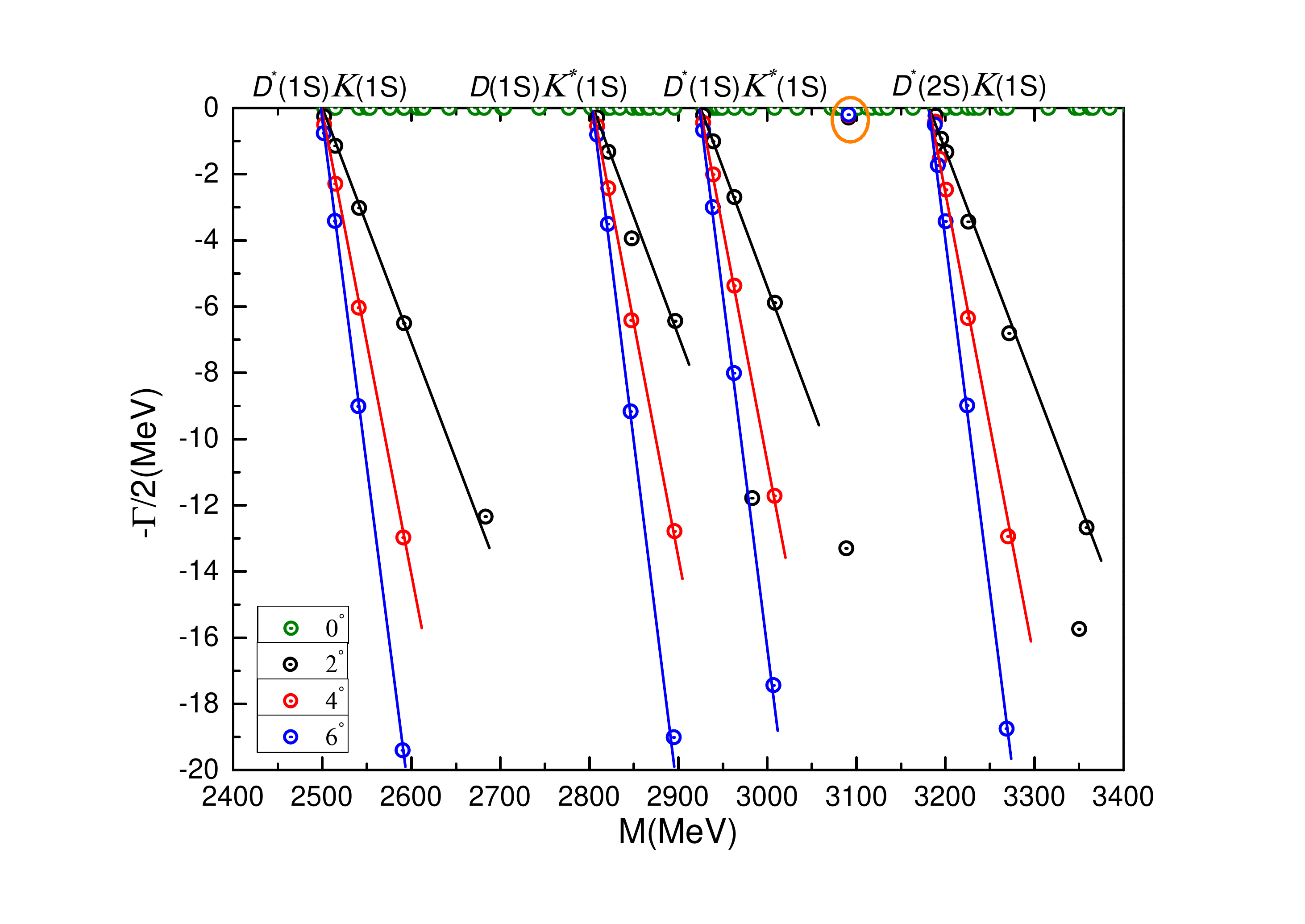}
\caption{\label{PP5} The complete coupled-channels calculation of the $sc\bar{q}\bar{q}$ tetraquark system with $I(J^P)=1(1^+)$ quantum numbers. We use the complex-scaling method of the chiral quark model varying $\theta$ from $0^\circ$ to $6^\circ$.}
\end{figure}

{\bf The $\bm{I(J^P)=1(1^+)}$ sector:} The numerical analysis of this case is as tedious as the $I(J^P)=0(1^+)$ because 30 channels must be still explored. From Table~\ref{GresultCC5}, we could conclude that no bound state is obtained when the three kinds of real-range calculations (single channel analysis, coupled-channels computation in each configuration of quarks and complete coupled-channels investigation) are performed. Theoretical masses of $D^*K$, $DK^*$ and $D^*K^*$ in color-singlet channels are located at 2498~MeV, 2804~MeV and 2924~MeV, respectively. The remaining channels, hidden-color, diquark-antidiquark and K-types, are generally at around 3.2 GeV, except for several channels of K-types that can be found at about 2.7 GeV and 3.7 GeV.

In a complex-range investigation of the complete coupled-channels calculation, where the angle $\theta$ is varied from $0^\circ$ to $6^\circ$, the energy dots are presented in Fig.~\ref{PP5}. Four scattering states, $D^*(1S)K(1S)$, $D(1S)K^*(1S)$, $D^*(1S)K^*(1S)$ and $D^*(2S)K(1S)$, are well shown within an energy interval 2.4-3.4 GeV. An extremely narrow resonance pole is found at 3091~MeV, its width is just 0.5~MeV. This state should also stable against two-mesons strong decays, and the dominant channel is $D^*(1S)K^*(1S)$. 

\begin{table}[!t]
\caption{\label{GresultCC6} Lowest-lying $sc\bar{q}\bar{q}$ tetraquark states with $I(J^P)=1(2^+)$ calculated within the real range formulation of the chiral quark model.
The allowed meson-meson, diquark-antidiquark and K-type configurations are listed in the first column; when possible, the experimental value of the non-interacting meson-meson threshold is labeled in parentheses. Each channel is assigned an index in the 2nd column, it reflects a particular combination of spin ($\chi_J^{\sigma_i}$), flavor ($\chi_I^{f_j}$) and color ($\chi_k^c$) wave functions that are shown explicitly in the 3rd column. The theoretical mass obatined in each channel is shown in the 4th column and the coupled result for each kind of configuration is presented in the 5th column.
When a complete coupled-channels calculation is performed, last row of the table indicates the lowest-lying mass and binding energy.
(unit: MeV).}
\begin{ruledtabular}
\begin{tabular}{lccc}
~~Channel   & Index & $\chi_J^{\sigma_i}$;~$\chi_I^{f_j}$;~$\chi_k^c$ & $M$~~ \\
        &   &$[i; ~j; ~k]$ &  \\[2ex]
$(D^* K^*)^1 (2899)$  & 1   & [1;~1;~3]  & $2924$  \\[2ex]
$(D^* K^*)^8$  & 2   & [1;~1;~3]   & $3139$   \\[2ex]
$(sc)^*(\bar{q}\bar{q})^*$  & 3   & [1;~1;~3]   & $3160$  \\[2ex]
$K_1$  & 4   & [1;~1;~3]   & $3149$  \\
            & 5   & [1;~1;~3]   & $3151$  \\[2ex]
$K_2$  & 6   & [1;~1;~3]   & $3057$  \\
             & 7   & [1;~1;~3]   & $3161$  \\[2ex]
$K_3$  & 8   & [1;~1;~3]   & $3782$  \\
             & 9   & [1;~1;~3]   & $3162$  \\[2ex]
$K_4$  & 10   & [1;~1;~3]   & $3156$  \\[2ex]
\multicolumn{3}{c}{Complete coupled-channels:} & $2924$
\end{tabular}
\end{ruledtabular}
\end{table}

\begin{figure}[ht]
\includegraphics[width=0.49\textwidth, trim={2.3cm 2.0cm 2.0cm 1.0cm}]{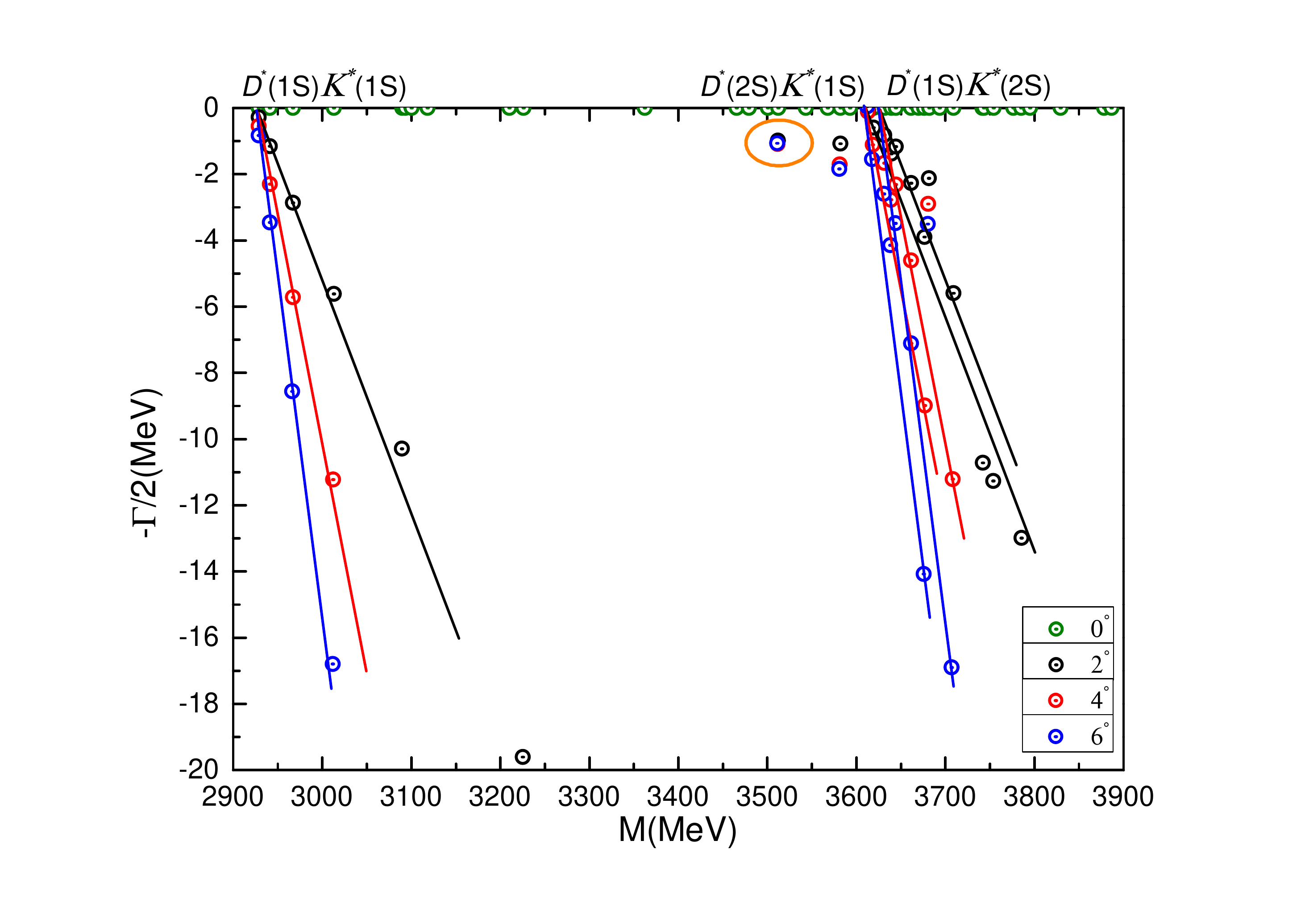}
\caption{\label{PP6} The complete coupled-channels calculation of the $sc\bar{q}\bar{q}$ tetraquark system with $I(J^P)=1(2^+)$ quantum numbers. We use the complex-scaling method of the chiral quark model varying $\theta$ from $0^\circ$ to $6^\circ$.}
\end{figure}

{\bf The $\bm{I(J^P)=1(2^+)}$ sector:} The real- and complex-range results for the highest spin and isospin $sc\bar{q}\bar{q}$ state are shown in Table~\ref{GresultCC6} and Fig.~\ref{PP6}, respectively.The only di-meson channel is the $D^*K^*$ one, and it is a scattering state with a theoretical mass of 2924~MeV. The states of hidden-color, diquark-antidiquark and K-type configurations are all above 3.0 GeV. Moreover, the scattering nature of the $D^*K^*$ channel remains in a complete coupled-channels calculation.

The $D^*(1S)K^*(1S)$ scattering state along with two almost degenerate radial excited states, $D^*(2S)K^*(1S)(3611)$ and $D^*(1S)K^*(2S)(3638)$, are well presented in Fig.~\ref{PP6}. Therein, a resonance pole is found, its mass and width are 3511~MeV and 2.1~MeV. Notice herein that a state of similar nature is found in the $0(2^+)$ $sc\bar{q}\bar{q}$ tetraquark sector.


\subsection{The $\mathbf{sb\bar{q}\bar{q}}$ tetraquarks}

A natural continuation of the investigation performed above is the analysis of the $sb\bar{q}\bar{q}$ bound-state problem with spin-parity $J^P=0^+$, $1^+$ and $2^+$, and isospin $I=0$ and $1$. A similar situation than the one discussed in the charm-strange sector is found for the bottom-strange tetraquark system. There are two bound states with quantum numbers $I(J^P)=0(0^+)$ and $0(1^+)$. Besides, resonances are also found in the former channels but also in the $I(J^P)=0(2^+)$, $1(0^+)$ and $1(2^+)$ ones. Further details can be found below.

\begin{table}[!t]
\caption{\label{GresultCC7} Lowest-lying $sb\bar{q}\bar{q}$ tetraquark states with $I(J^P)=0(0^+)$ calculated within the real range formulation of the chiral quark model.
The allowed meson-meson, diquark-antidiquark and K-type configurations are listed in the first column; when possible, the experimental value of the non-interacting meson-meson threshold is labeled in parentheses. Each channel is assigned an index in the 2nd column, it reflects a particular combination of spin ($\chi_J^{\sigma_i}$), flavor ($\chi_I^{f_j}$) and color ($\chi_k^c$) wave functions that are shown explicitly in the 3rd column. The theoretical mass obatined in each channel is shown in the 4th column and the coupled result for each kind of configuration is presented in the 5th column.
When a complete coupled-channels calculation is performed, last row of the table indicates the lowest-lying mass and binding energy.
(unit: MeV).}
\begin{ruledtabular}
\begin{tabular}{lcccc}
~~Channel   & Index & $\chi_J^{\sigma_i}$;~$\chi_I^{f_j}$;~$\chi_k^c$ & $M$ & Mixed~~ \\
        &   &$[i; ~j; ~k]$ &  \\[2ex]
$(B K)^1 (5774)$          & 1  & [1;~1;~3]  & $5755$ & \\
$(B^* K^*)^1 (6217)$  & 2  & [2;~1;~3]   & $6223$ & $5752$ \\[2ex]
$(B K)^8$      & 3  & [1;~1;~4]   & $6519$ & \\
$(B^* K^*)^8$  & 4  & [2;~1;~4]   & $6378$ & $6217$ \\[2ex]
$(sb)(\bar{q}\bar{q})$      & 5   & [3;~1;~1]  & $6044$ & \\
$(sb)^*(\bar{q}\bar{q})^*$  & 6  & [4;~1;~2]   & $6367$ & $5965$ \\[2ex]
$K_1$  & 7  & [5;~1;~5]   & $6228$ & \\
  & 8  & [5;~1;~6]   & $6341$ & \\
  & 9  & [6;~1;~5]   & $6475$ & \\
  & 10  & [6;~1;~6]   & $5957$ & $5890$ \\[2ex]
$K_2$  & 11  & [7;~1;~7]   & $6235$ & \\
  & 12  & [7;~1;~8]   & $6375$ & \\
  & 13  & [8;~1;~7]   & $5784$ & \\
  & 14  & [8;~1;~8]   & $6511$ & $5773$ \\[2ex]
$K_3$  & 15  & [9;~1;~9]   & $6351$ & \\
  & 16  & [9;~1;~10]   & $7049$ & \\
  & 17  & [10;~1;~9]   & $7093$ & \\
  & 18  & [10;~1;~10]   & $6185$ & $6076$ \\[2ex]
$K_4$  & 19  & [11;~1;~11]   & $6306$ & \\
  & 20  & [12;~1;~12]   & $6036$ & $5925$ \\[2ex]
\multicolumn{4}{c}{Complete coupled-channels:} & $5710$ \\
   &   &    &  & $E_B=-49$
\end{tabular}
\end{ruledtabular}
\end{table}

\begin{table}[!t]
\caption{\label{GresultComp3} Relative strengths of various meson-meson, diquark-antidiquark and K-type components in the wave function of the $I(J^P)=0(0^+)$ $sb\bar{q}\bar{q}$ tetraquark bound-state obtained when a complete coupled-channels calculation is performed. The superscripts $1$ and $8$ are for singlet-color and hidden-color channels, respectively.}
\begin{ruledtabular}
\begin{tabular}{lcccc}
  ~~$(BK)^1$~~  & ~~$(B^*K^*)^1$~~   & ~~$(BK)^8$~~ &
   ~~$(B^*K^*)^8$~~ & ~~$(sb)(\bar{q}\bar{q})$~~ \\
 ~~22.7\%~~  & ~~18.0\%~~  & ~~1.4\%  ~~& 4.0\% ~~ & 5.6\%~~\\[2ex]
 ~~$(sb)^*(\bar{q}\bar{q})^*$~~ & ~~$K_1$~~ & ~~$K_2$~~ & ~~$K_3$~~ & ~~$K_4$~~ \\ 
 ~~2.3\%~~  & ~~9.9\%~~  & ~~20.9\%  ~~& 12.7\% ~~ & 2.5\%~~\\
\end{tabular}
\end{ruledtabular}
\end{table}

\begin{table}[!t]
\caption{\label{tab:dis3} The distance, in fm, between any two quarks of the $I(J^P)=0(0^+)$ $sb\bar{q}\bar{q}$ tetraquark bound-state obtained when a complete coupled-channels calculation is performed.}
\begin{ruledtabular}
\begin{tabular}{cccc}
  ~~$r_{\bar{q}\bar{q}}$ & $r_{\bar{q}s}$ & $r_{\bar{q}b}$ & $r_{sb}$~~  \\[2ex]
  ~~0.97 & 0.78 & 0.75 & 0.85~~ \\
\end{tabular}
\end{ruledtabular}
\end{table}

\begin{figure}[ht]
\includegraphics[width=0.49\textwidth, trim={2.3cm 2.0cm 2.0cm 1.0cm}]{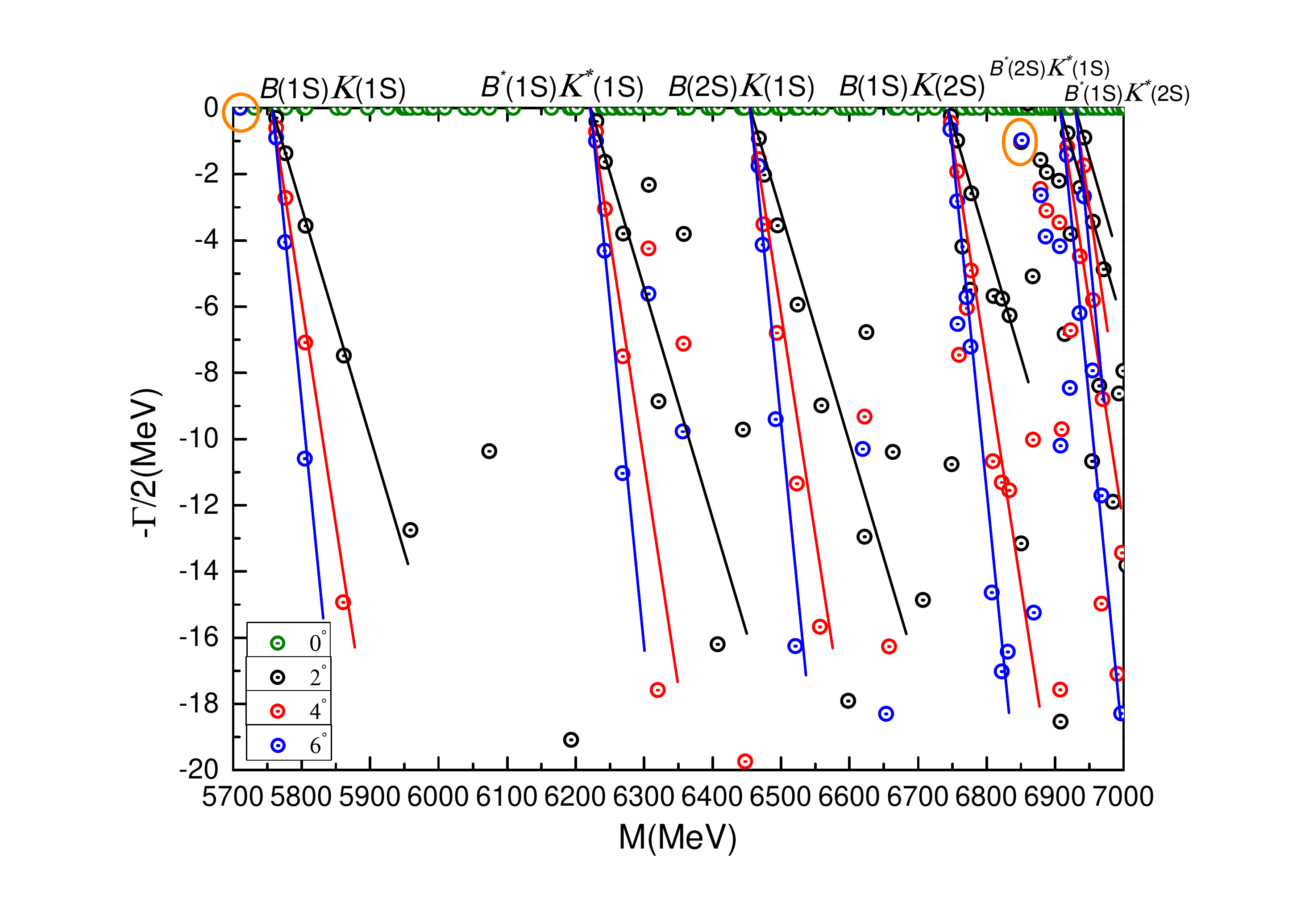}
\caption{\label{PP7} The complete coupled-channels calculation of the $sb\bar{q}\bar{q}$ tetraquark system with $I(J^P)=0(0^+)$ quantum numbers. We use the complex-scaling method of the chiral quark model varying $\theta$ from $0^\circ$ to $6^\circ$.}
\end{figure}

{\bf The $\bm{I(J^P)=0(0^+)}$ sector:} Table~\ref{GresultCC7} summarizes all contributing channels, they are 20 and include meson-meson, diquark-antidiquark and K-type structures. A loosely bound state of $BK$ is obtained in the color-singlet channel calculation, the predicted mass is 5755~MeV, which means a binding energy of $-2$~MeV. The color-singlet $B^*K^*$ channel is a scattering state with a mass just located at the value of theoretical threshold, 6223~MeV. Besides, the remaining channels of diquark-antidiquark and K-type configurations are generally located in the interval 6.0-7.0 GeV.

If a coupled-channels calculation is performed in each individual quark configuration, the bound state of $BK$ is further pushed down and its coupled-mass is 5752~MeV. When all 20 channels listed in Table~\ref{GresultCC7} are considered in a coupled-channels calculation in real-range, the lowest mass at 5710~MeV indicates a binding energy of $-49$~MeV. Therefore, a strong coupling effect is confirmed in the $I(J^P)=0(0^+)$ $sc\bar q\bar q$ tetraquark sector, as it is the case in the $sc\bar{q}\bar{q}$ one. We have investigated the composition of this state in Table~\ref{GresultComp3}. The dominant structure is of K-type contributing 44\% to the total wave function, but singlet-color $BK$ (23\%) and $B^*K^*$ (18\%) components are, when combined, equally important. Table~\ref{tab:dis3} indicates that all quark--(anti-)quark pairs are separated about the same distance with a value below 1 fm. Our interpretation is that this state looks like a compact tetraquark structure.

Figure~\ref{PP7} shows the distribution of complex energies when the complete coupled-channels calculation is extended to the complex plane. Six scattering states, \emph{i.e.} $B(1S)K(1S)$, $B^*(1S)K^*(1S)$, $B(2S)K(1S)$, $B(1S)K(2S)$, $B^*(2S)K^*(1S)$ and $B^*(1S)K^*(2S)$, are well identified in the energy interval 5.7-7.0 GeV; there is a degeneration between  $B^*(2S)K^*(1S)(6912)$ and $B^*(1S)K^*(2S)(6940)$. Two singularities deserve to be highlighted in Fig.~\ref{PP7}, the bound state at a real-axis fixed mass of 5710~MeV and a narrow resonance whose parameters are $M=6850$~MeV and $\Gamma=2.0$~MeV. The last one is independent of the rotated angle $\theta$, which is varied from $0^\circ$ to $6^\circ$, and it can be identified as $B(1S)K(2S)$ resonance state because its location is far from $B(1S)K(1S)$, $B^*(1S)K^*(1S)$ and $B(2S)K(1S)$ threshold lines.

\begin{table}[!t]
\caption{\label{GresultCC8} Lowest-lying $sb\bar{q}\bar{q}$ tetraquark states with $I(J^P)=0(1^+)$ calculated within the real range formulation of the chiral quark model.
The allowed meson-meson, diquark-antidiquark and K-type configurations are listed in the first column; when possible, the experimental value of the non-interacting meson-meson threshold is labeled in parentheses. Each channel is assigned an index in the 2nd column, it reflects a particular combination of spin ($\chi_J^{\sigma_i}$), flavor ($\chi_I^{f_j}$) and color ($\chi_k^c$) wave functions that are shown explicitly in the 3rd column. The theoretical mass obatined in each channel is shown in the 4th column and the coupled result for each kind of configuration is presented in the 5th column.
When a complete coupled-channels calculation is performed, last row of the table indicates the lowest-lying mass and binding energy.
(unit: MeV).}
\scalebox{0.949}{\begin{ruledtabular}
\begin{tabular}{lcccc}
~~Channel   & Index & $\chi_J^{\sigma_i}$;~$\chi_I^{f_j}$;~$\chi_k^c$ & $M$ & Mixed~~ \\
        &   &$[i; ~j; ~k]$ &  \\[2ex]
$(B^* K)^1 (5819)$      & 1    & [1;~1;~3]   & $5797$ & \\
$(B K^*)^1 (6172)$      & 2    & [2;~1;~3]    & $6185$ & \\
$(B^* K^*)^1 (6217)$  & 3    & [3;~1;~3]    & $6226$ & $5795$ \\[2ex]
$(B^* K)^8$          & 4    & [1;~1;~4]    & $6530$ & \\
$(B K^*)^8$          & 5    & [2;~1;~4]    & $6587$ & \\
$(B^* K^*)^8$     & 6    & [3;~1;~4]    & $6547$ & $6246$ \\[2ex]
$(sb)(\bar{q}\bar{q})^*$   & 7    & [4;~1;~1]    & $6057$ &  \\
$(sb)^*(\bar{q}\bar{q})$   & 8    & [5;~1;~2]    & $6511$ &  \\
$(sb)^*(\bar{q}\bar{q})^*$   & 9    & [6;~1;~2]    & $6419$ & $5989$ \\[2ex]
$K_1$  & 10    & [7;~1;~5]    & $6588$ & \\
  & 11    & [7;~1;~6]    & $6471$ & \\
  & 12    & [8;~1;~5]    & $6256$ & \\
  & 13    & [8;~1;~6]    & $6341$ & \\
  & 14    & [9;~1;~5]    & $6483$ & \\
  & 15    & [9;~1;~6]    & $5980$ & $5919$ \\[2ex]
$K_2$  & 16    & [10;~1;~7]    & $6221$ & \\
  & 17    & [10;~1;~8]    & $6470$ & \\
  & 18    & [11;~1;~7]    & $6255$ & \\
  & 19    & [11;~1;~8]    & $6592$ & \\
  & 20    & [12;~1;~7]    & $5820$ & \\
  & 21    & [12;~1;~8]    & $6522$ & $5812$ \\[2ex]
$K_3$  & 22    & [13;~1;~9]    & $6413$ & \\
  & 23    & [13;~1;~10]    & $6252$ & \\
  & 24    & [14;~1;~9]    & $6412$ & \\
  & 25    & [14;~1;~10]    & $6254$ & \\
  & 26    & [15;~1;~9]    & $6501$ & \\
  & 27    & [15;~1;~10]    & $7092$ & $6109$ \\[2ex]
$K_4$  & 28    & [16;~1;~11]    & $6499$ & \\
  & 29    & [17;~1;~11]    & $6325$ & \\
  & 30    & [18;~1;~12]    & $6046$ & $5947$ \\[2ex]
\multicolumn{4}{c}{Complete coupled-channels:} & $5754$\\
   &   &    &  & $E_B=-46$
\end{tabular}
\end{ruledtabular}}
\end{table}

\begin{table}[!t]
\caption{\label{GresultComp4} Relative strengths of various meson-meson, diquark-antidiquark and K-type components in the wave function of the $I(J^P)=0(0^+)$ $sb\bar{q}\bar{q}$ tetraquark bound-state obtained when a complete coupled-channels calculation is performed. The superscripts $1$ and $8$ are for singlet-color and hidden-color channels, respectively.}
\begin{ruledtabular}
\begin{tabular}{ccccc}
  ~~$(B^*K)^1$~~  & ~~$(BK^*)^1$~~   & ~~$(B^*K^*)^1$~~ &
  ~~$(B^*K)^8$~~  & ~~$(BK^*)^8$~~  \\
 ~~4.4\%~~  & ~~2.7\%~~  & ~~3.0\%  ~~& 0.4\% ~~ & 0.5\%~~\\[2ex]
  ~~$(B^*K^*)^8$~~ & ~~$(sc)(\bar{q}\bar{q})^*$~~  & ~~$(sc)^*(\bar{q}\bar{q})$~~   & ~~$(sc)^*(\bar{q}\bar{q})^*$~~ & ~~$K_1$~~ \\
 ~~0.7\%~~  & ~~1.1\%~~  & ~~0.2\%  ~~& 0.2\% ~~ & 2.6\%~~\\[2ex]
 ~~$K_2$~~ & ~~$K_3$~~ & ~~$K_4$~~ &  &  \\ 
 ~~5.6\%~~  & ~~78.0\%~~  & ~~0.5\%  ~~&  & \\
\end{tabular}
\end{ruledtabular}
\end{table}

\begin{table}[!t]
\caption{\label{tab:dis4} The distance, in fm, between any two quarks of the $I(J^P)=0(1^+)$ $sb\bar{q}\bar{q}$ tetraquark bound-state obtained when a complete coupled-channels calculation is performed.}
\begin{ruledtabular}
\begin{tabular}{cccc}
  ~~$r_{\bar{q}\bar{q}}$ & $r_{\bar{q}s}$ & $r_{\bar{q}b}$ & $r_{sb}$~~  \\[2ex]
  ~~0.99 & 0.79 & 0.77 & 0.87~~ \\
\end{tabular}
\end{ruledtabular}
\end{table}

\begin{figure}[ht]
\includegraphics[width=0.49\textwidth, trim={2.3cm 2.0cm 2.0cm 1.0cm}]{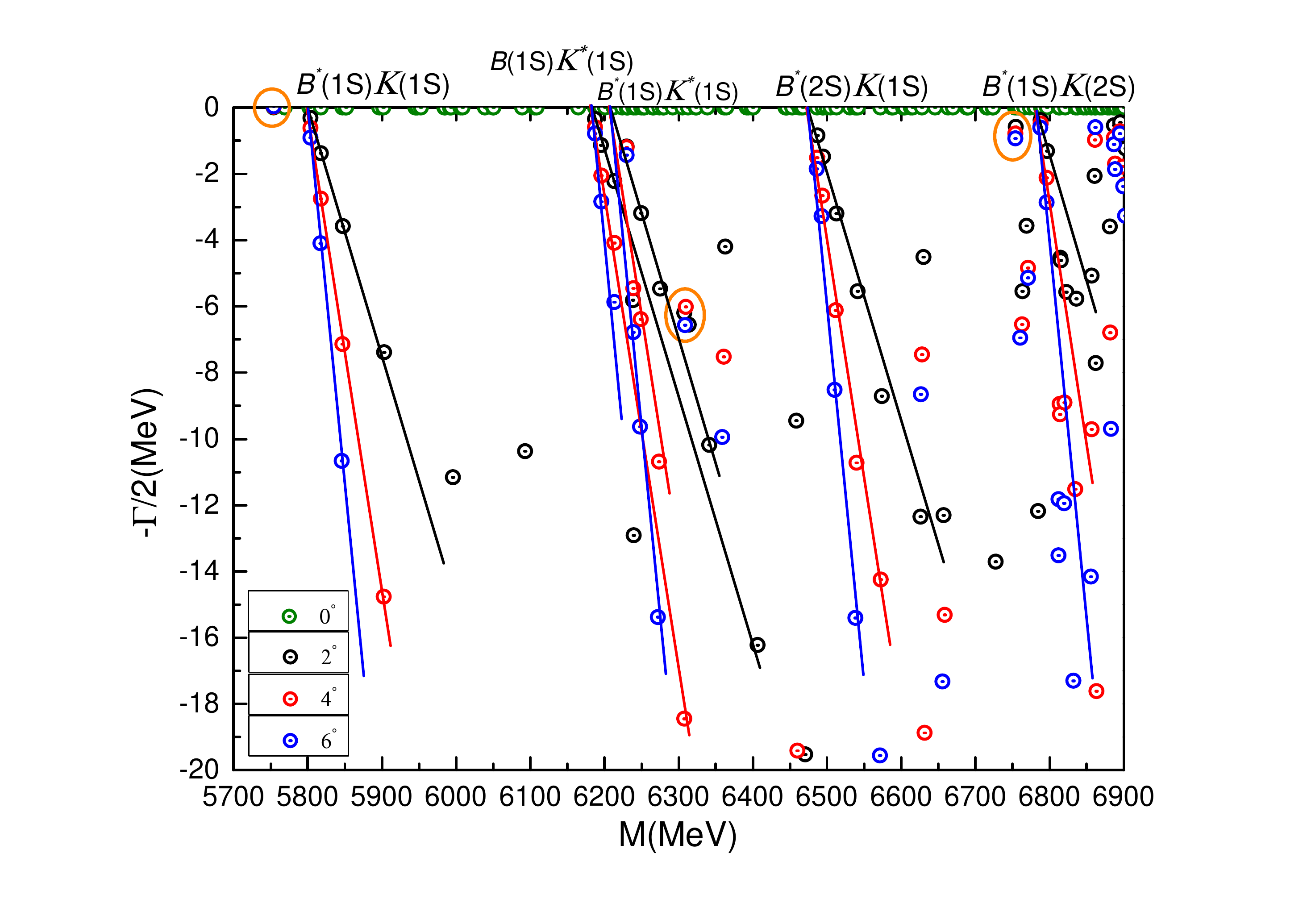}
\caption{\label{PP8} The complete coupled-channels calculation of the $sb\bar{q}\bar{q}$ tetraquark system with $I(J^P)=0(1^+)$ quantum numbers. We use the complex-scaling method of the chiral quark model varying $\theta$ from $0^\circ$ to $6^\circ$.}
\end{figure}

{\bf The $\bm{I(J^P)=0(1^+)}$ sector:} 30 channels shown in Table~\ref{GresultCC8} are under investigation; namely, three meson-meson channels in both color-singlet and hidden-color configurations, three diquark-antidiquark structures, and 21 K-type arrangements. In the single-channel calculation, we obtain a bound state with binding energy $-3$~MeV and total mass 5797~MeV. The other channels are generally above 6.0 GeV, except for two cases, a $K_1$ channel (5980~MeV) and a $K_2$ one (5820~MeV).

The coupled-channels calculation of color-singlet meson-meson structures delivers a slightly deeper bound state with a mass 5795~MeV and $E_B=-5$~MeV. When the complete coupled-channels computation is performed in real-range the binding energy changes to be $-46$~MeV, and thus a deeply bound state with mass equal to 5754~MeV is obtained. It appears that the increase in binding energy is due to a strong coupling effect between channels. We have investigated the composition of this state in Table~\ref{GresultComp4}. The dominant structure is of K-type contributing 74\% to the total wave function, but traces of singlet-color $B^{(*)}K^{(*)}$ (12\%) can be found. Table~\ref{tab:dis4} indicates that all quark--(anti-)quark pairs are separated about the same distance with a value below 1 fm. Our interpretation is that this state looks like a compact tetraquark structure.

Figure~\ref{PP8} shows our findings using the complex-range method to the complete coupled channels calculation. The scattering states of $B^*(1S)K(1S)$, $B(1S)K^*(1S)$, $B^*(1S)K^*(1S)$, $B^*(2S)K(1S)$ and $B^*(1S)K(2S)$ are clearly identified in the energy region 5.7-6.9 GeV, and there are two almost degenerate states: $B(1S)K^*(1S)(6185)$ and $B^*(1S)K^*(1S)(6226)$. The tightly bound state at 5754~MeV is independent of the rotated angle, which is varied from $0^\circ$ to $6^\circ$. Two resonances are found whose parameters are $(6313,12.5)$~MeV and $(6754,1.2)$~MeV, respectively; they can be identified as $B^*(1S)K^*(1S)$ and $B^*(2S)K(1S)$ resonances.

\begin{table}[!t]
\caption{\label{GresultCC9} Lowest-lying $sb\bar{q}\bar{q}$ tetraquark states with $I(J^P)=0(2^+)$ calculated within the real range formulation of the chiral quark model.
The allowed meson-meson, diquark-antidiquark and K-type configurations are listed in the first column; when possible, the experimental value of the non-interacting meson-meson threshold is labeled in parentheses. Each channel is assigned an index in the 2nd column, it reflects a particular combination of spin ($\chi_J^{\sigma_i}$), flavor ($\chi_I^{f_j}$) and color ($\chi_k^c$) wave functions that are shown explicitly in the 3rd column. The theoretical mass obatined in each channel is shown in the 4th column and the coupled result for each kind of configuration is presented in the 5th column.
When a complete coupled-channels calculation is performed, last row of the table indicates the lowest-lying mass and binding energy.
(unit: MeV).}
\begin{ruledtabular}
\begin{tabular}{lccc}
~~Channel   & Index & $\chi_J^{\sigma_i}$;~$\chi_I^{f_j}$;~$\chi_k^c$ & $M$~~ \\
        &   &$[i; ~j; ~k]$ &  \\[2ex]
$(B^* K^*)^1 (6217)$  & 1   & [1;~1;~3]  & $6226$  \\[2ex]
$(B^* K^*)^8$  & 2   & [1;~1;~3]   & $6611$   \\[2ex]
$(sb)^*(\bar{q}\bar{q})^*$  & 3   & [1;~1;~3]   & $6505$  \\[2ex]
$K_1$  & 4   & [1;~1;~3]   & $6596$  \\
            & 5   & [1;~1;~3]   & $6481$  \\[2ex]
$K_2$  & 6   & [1;~1;~3]   & $6267$  \\
             & 7   & [1;~1;~3]   & $6604$  \\[2ex]
$K_3$  & 8   & [1;~1;~3]   & $6497$  \\
             & 9   & [1;~1;~3]   & $7086$  \\[2ex]
$K_4$  & 10   & [1;~1;~3]   & $6510$  \\[2ex]
\multicolumn{3}{c}{Complete coupled-channels:} & $6226$
\end{tabular}
\end{ruledtabular}
\end{table}

\begin{figure}[ht]
\includegraphics[width=0.49\textwidth, trim={2.3cm 2.0cm 2.0cm 1.0cm}]{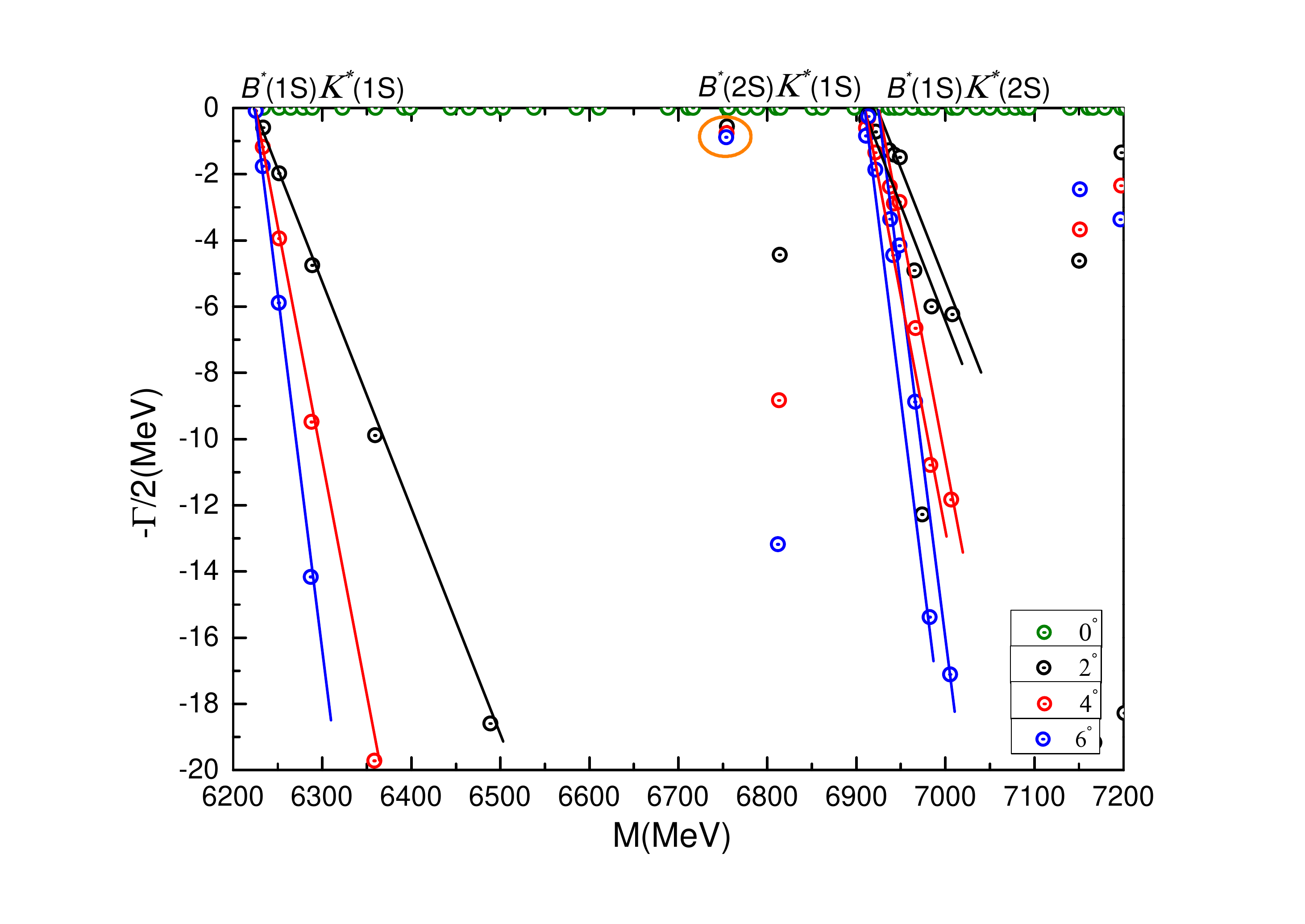}
\caption{\label{PP9} The complete coupled-channels calculation of the $sb\bar{q}\bar{q}$ tetraquark system with $I(J^P)=0(2^+)$ quantum numbers. We use the complex-scaling method of the chiral quark model varying $\theta$ from $0^\circ$ to $6^\circ$.}
\end{figure}

{\bf The $\bm{I(J^P)=0(2^+)}$ sector:} From Table~\ref{GresultCC9}, we can see that the 10 contributing channels are all above 6.2 GeV and the lowest mass is 6226~MeV, which is just the theoretical value of $B^*K^*$ threshold. Moreover, this fact is not changed when a complete coupled-channels calculation is performed in real-range.

In the complex-scaling investigation, the scattering states of $B^*(1S)K^*(1S)$, $B^*(2S)K^*(1S)$ and $B^*(1S)K^*(2S)$ are well identified when the rotated angle is varied from $0^\circ$ to $6^\circ$. They are located within an energy interval 6.2-7.2 GeV (see Fig.~\ref{PP9}). It is worth highlighting that a pole is obtained between the threshold lines of  $B^*(1S)K^*(1S)$ and the two almost degenerate $B^*(2S)K^*(1S)(6912)$ and $B^*(1S)K^*(2S)(6940)$. Accordingly, it can be identified as a $B^*(1S)K^*(1S)$ resonance with mass and width 6754~MeV and 1.2~MeV, respectively.


\begin{table}[!t]
\caption{\label{GresultCC10} Lowest-lying $sb\bar{q}\bar{q}$ tetraquark states with $I(J^P)=1(0^+)$ calculated within the real range formulation of the chiral quark model.
The allowed meson-meson, diquark-antidiquark and K-type configurations are listed in the first column; when possible, the experimental value of the non-interacting meson-meson threshold is labeled in parentheses. Each channel is assigned an index in the 2nd column, it reflects a particular combination of spin ($\chi_J^{\sigma_i}$), flavor ($\chi_I^{f_j}$) and color ($\chi_k^c$) wave functions that are shown explicitly in the 3rd column. The theoretical mass obatined in each channel is shown in the 4th column and the coupled result for each kind of configuration is presented in the 5th column.
When a complete coupled-channels calculation is performed, last row of the table indicates the lowest-lying mass and binding energy.
(unit: MeV).}
\begin{ruledtabular}
\begin{tabular}{lcccc}
~~Channel   & Index & $\chi_J^{\sigma_i}$;~$\chi_I^{f_j}$;~$\chi_k^c$ & $M$ & Mixed~~ \\
        &   &$[i; ~j; ~k]$ &  \\[2ex]
$(B K)^1 (5774)$          & 1  & [1;~1;~3]  & $5759$ & \\
$(B^* K^*)^1 (6217)$  & 2  & [2;~1;~3]   & $6226$ & $5759$ \\[2ex]
$(B K)^8$      & 3  & [1;~1;~4]   & $6622$ & \\
$(B^* K^*)^8$  & 4  & [2;~1;~4]   & $6606$ & $6353$ \\[2ex]
$(sb)(\bar{q}\bar{q})$      & 5   & [3;~1;~1]  & $6548$ & \\
$(sb)^*(\bar{q}\bar{q})^*$  & 6  & [4;~1;~2]   & $6390$ & $6362$ \\[2ex]
$K_1$  & 7  & [5;~1;~5]   & $6563$ & \\
  & 8  & [5;~1;~6]   & $6514$ & \\
  & 9  & [6;~1;~5]   & $6581$ & \\
  & 10  & [6;~1;~6]   & $6096$ & $6084$ \\[2ex]
$K_2$  & 11  & [7;~1;~7]   & $6290$ & \\
  & 12  & [7;~1;~8]   & $6598$ & \\
  & 13  & [8;~1;~7]   & $5850$ & \\
  & 14  & [8;~1;~8]   & $6625$ & $5848$ \\[2ex]
$K_3$  & 15  & [9;~1;~9]   & $7011$ & \\
  & 16  & [9;~1;~10]   & $6390$ & \\
  & 17  & [10;~1;~9]   & $6540$ & \\
  & 18  & [10;~1;~10]   & $7085$ & $6340$ \\[2ex]
$K_4$  & 19  & [11;~1;~11]   & $6365$ & \\
  & 20  & [12;~1;~12]   & $6551$ & $6321$ \\[2ex]
\multicolumn{4}{c}{Complete coupled-channels:} & $5759$
\end{tabular}
\end{ruledtabular}
\end{table}

\begin{figure}[ht]
\includegraphics[width=0.49\textwidth, trim={2.3cm 2.0cm 2.0cm 1.0cm}]{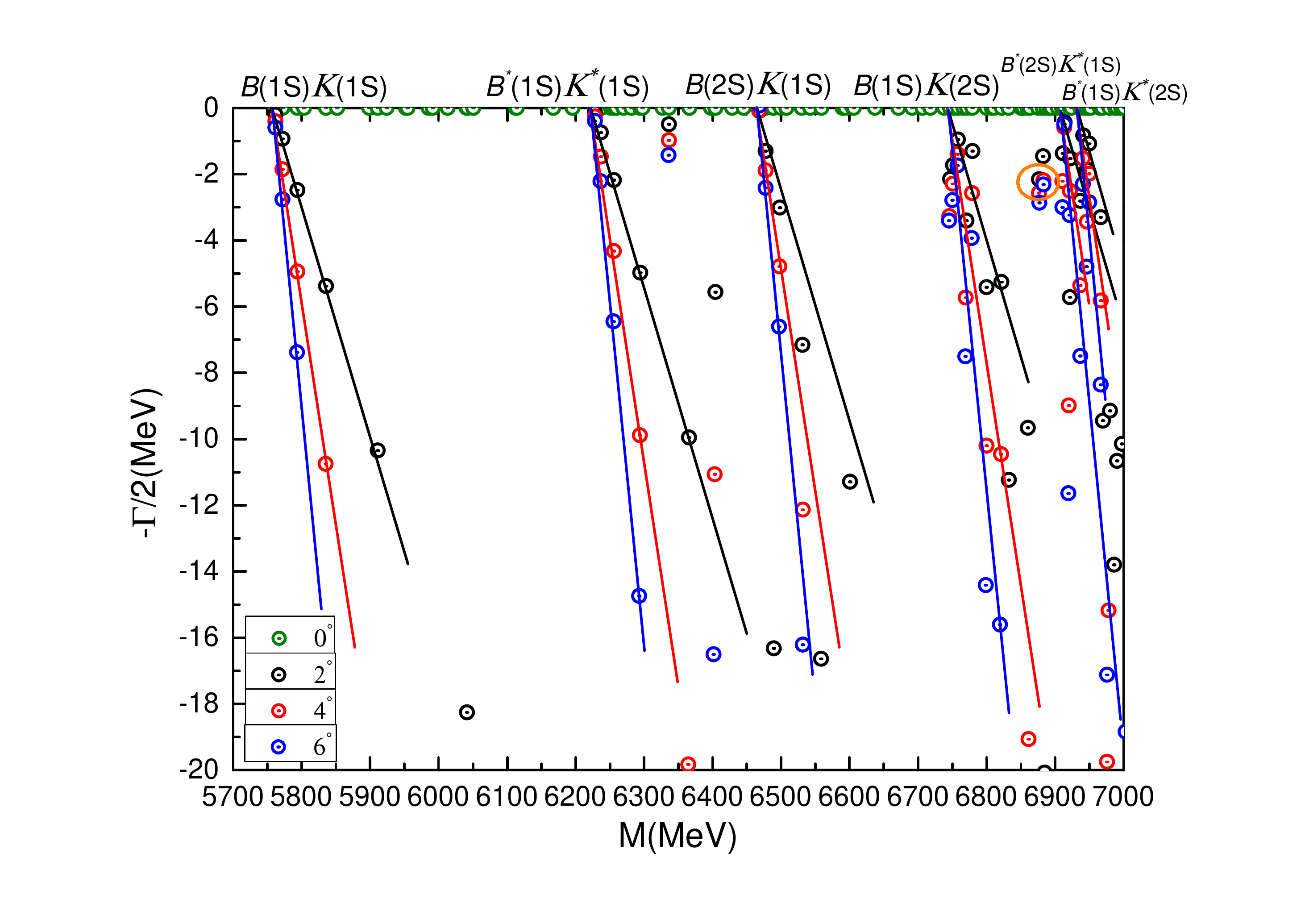}
\caption{\label{PP10} The complete coupled-channels calculation of the $sb\bar{q}\bar{q}$ tetraquark system with $I(J^P)=1(0^+)$ quantum numbers. We use the complex-scaling method of the chiral quark model varying $\theta$ from $0^\circ$ to $6^\circ$.}
\end{figure}

{\bf The $\bm{I(J^P)=1(0^+)}$ sector:} Table~\ref{GresultCC10} lists the calculated masses of the $1(0^+)$ $sb\bar{q}\bar{q}$ tetraquark state. As in the charm-strange sector, no bound state is found in each single channel and in the different variants of the coupled-channels calculations. The lowest mass obtained is 5759~MeV, which is just the theoretical value of the $BK$ threshold. The remaining channels are generally located within the energy interval 6.1-7.0 GeV, except for one $K_2$ channel which presents a mass of 5850~MeV.

When the complete coupled-channels calculation is performed using the complex-range method, the scattering nature of $B(1S)K(1S)$, $B^*(1S)K^*(1S)$, $B(2S)K(1S)$, $B(1S)K(2S)$, $B^*(2S)K^*(1S)$ and $B^*(1S)K^*(2S)$ are clearly shown in Fig.~\ref{PP10}. There is a dense distribution of complex energies in a region between 6.9 GeV and 7.0 GeV. One fixed resonance pole obviously emerges therein; its resonance parameters are $M=6876$~MeV and $\Gamma=4.2$~MeV. The dominant channel of this narrow resonance state is $B(1S)K(2S)$.

\begin{table}[!t]
\caption{\label{GresultCC11} Lowest-lying $sb\bar{q}\bar{q}$ tetraquark states with $I(J^P)=1(1^+)$ calculated within the real range formulation of the chiral quark model.
The allowed meson-meson, diquark-antidiquark and K-type configurations are listed in the first column; when possible, the experimental value of the non-interacting meson-meson threshold is labeled in parentheses. Each channel is assigned an index in the 2nd column, it reflects a particular combination of spin ($\chi_J^{\sigma_i}$), flavor ($\chi_I^{f_j}$) and color ($\chi_k^c$) wave functions that are shown explicitly in the 3rd column. The theoretical mass obatined in each channel is shown in the 4th column and the coupled result for each kind of configuration is presented in the 5th column.
When a complete coupled-channels calculation is performed, last row of the table indicates the lowest-lying mass and binding energy.
(unit: MeV).}
\scalebox{0.97}{\begin{ruledtabular}
\begin{tabular}{lcccc}
~~Channel   & Index & $\chi_J^{\sigma_i}$;~$\chi_I^{f_j}$;~$\chi_k^c$ & $M$ & Mixed~~ \\
        &   &$[i; ~j; ~k]$ &  \\[2ex]
$(B^* K)^1 (5819)$      & 1    & [1;~1;~3]   & $5800$ & \\
$(B K^*)^1 (6172)$      & 2    & [2;~1;~3]    & $6185$ & \\
$(B^* K^*)^1 (6217)$  & 3    & [3;~1;~3]    & $6226$ & $5800$ \\[2ex]
$(B^* K)^8$          & 4    & [1;~1;~4]    & $6611$ & \\
$(B K^*)^8$          & 5    & [2;~1;~4]    & $6583$ & \\
$(B^* K^*)^8$     & 6    & [3;~1;~4]    & $6601$ & $6363$ \\[2ex]
$(sb)(\bar{q}\bar{q})^*$   & 7    & [4;~1;~1]    & $6546$ &  \\
$(sb)^*(\bar{q}\bar{q})$   & 8    & [5;~1;~2]    & $6465$ &  \\
$(sb)^*(\bar{q}\bar{q})^*$   & 9    & [6;~1;~2]    & $6411$ & $6374$ \\[2ex]
$K_1$  & 10    & [7;~1;~5]    & $6426$ & \\
  & 11    & [7;~1;~6]    & $6428$ & \\
  & 12    & [8;~1;~5]    & $6564$ & \\
  & 13    & [8;~1;~6]    & $6507$ & \\
  & 14    & [9;~1;~5]    & $6573$ & \\
  & 15    & [9;~1;~6]    & $6104$ & $6092$ \\[2ex]
$K_2$  & 16    & [10;~1;~7]    & $6273$ & \\
  & 17    & [10;~1;~8]    & $6603$ & \\
  & 18    & [11;~1;~7]    & $6290$ & \\
  & 19    & [11;~1;~8]    & $6507$ & \\
  & 20    & [12;~1;~7]    & $5877$ & \\
  & 21    & [12;~1;~8]    & $6614$ & $5876$ \\[2ex]
$K_3$  & 22    & [13;~1;~9]    & $6517$ & \\
  & 23    & [13;~1;~10]    & $6434$ & \\
  & 24    & [14;~1;~9]    & $6515$ & \\
  & 25    & [14;~1;~10]    & $6434$ & \\
  & 26    & [15;~1;~9]    & $7096$ & \\
  & 27    & [15;~1;~10]    & $6469$ & $6360$ \\[2ex]
$K_4$  & 28    & [16;~1;~11]    & $6429$ & \\
  & 29    & [17;~1;~11]    & $6369$ & \\
  & 30    & [18;~1;~12]    & $6548$ & $6333$ \\[2ex]
\multicolumn{4}{c}{Complete coupled-channels:} & $5800$
\end{tabular}
\end{ruledtabular}}
\end{table}

\begin{figure}[ht]
\includegraphics[width=0.49\textwidth, trim={2.3cm 2.0cm 2.0cm 1.0cm}]{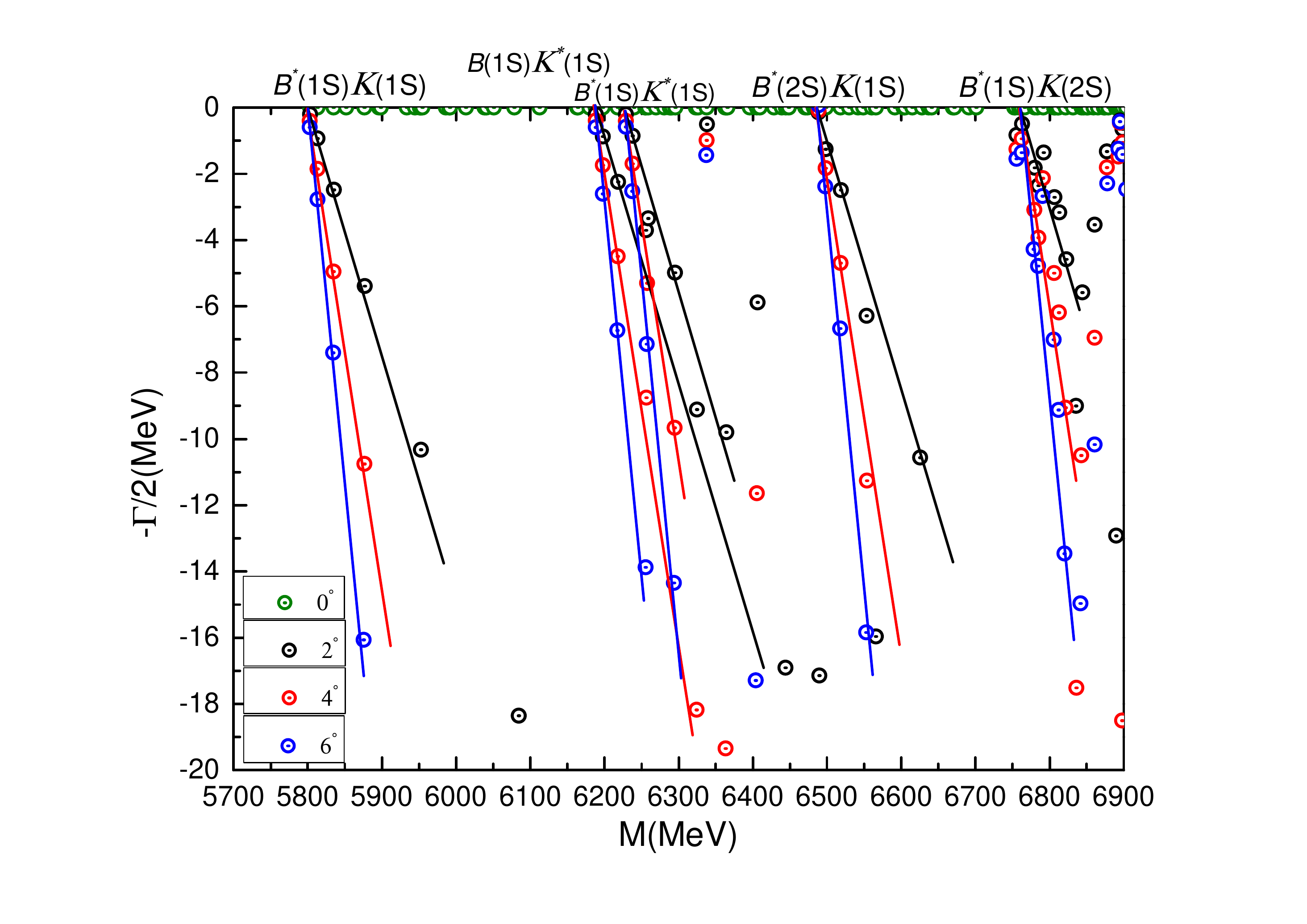}
\caption{\label{PP11} The complete coupled-channels calculation of the $sb\bar{q}\bar{q}$ tetraquark system with $I(J^P)=1(1^+)$ quantum numbers. We use the complex-scaling method of the chiral quark model varying $\theta$ from $0^\circ$ to $6^\circ$.}
\end{figure}

{\bf The $\bm{I(J^P)=1(1^+)}$ sector:} 30 channels contribute to the $sb\bar{q}\bar{q}$ tetraquark system with $1(1^+)$ quantum numbers.  The three meson-meson channels, $B^*K$, $BK^*$ and $B^*K^*$, are all scattering states and the lowest mass 5.8 GeV is just the value of non-interacting $B^*K$ threshold. The other exotic quark arrangements are generally above 6.2 GeV, except for one $K_2$ channel at 5.88 GeV. In a further step, when coupled-channels calculations are performed within real- and complex-range, neither bound nor resonance states are available. In particular, five scattering states, $B^*(1S)K(1S)$, $B(1S)K^*(1S)$, $B^*(1S)K^*(1S)$, $B^*(2S)K(1S)$ and $B^*(1S)K(2S)$, are presented in Fig.~\ref{PP11} and one can see that bound and resonance poles do not emerge.

\begin{table}[!t]
\caption{\label{GresultCC12} Lowest-lying $sb\bar{q}\bar{q}$ tetraquark states with $I(J^P)=1(2^+)$ calculated within the real range formulation of the chiral quark model.
The allowed meson-meson, diquark-antidiquark and K-type configurations are listed in the first column; when possible, the experimental value of the non-interacting meson-meson threshold is labeled in parentheses. Each channel is assigned an index in the 2nd column, it reflects a particular combination of spin ($\chi_J^{\sigma_i}$), flavor ($\chi_I^{f_j}$) and color ($\chi_k^c$) wave functions that are shown explicitly in the 3rd column. The theoretical mass obatined in each channel is shown in the 4th column and the coupled result for each kind of configuration is presented in the 5th column.
When a complete coupled-channels calculation is performed, last row of the table indicates the lowest-lying mass and binding energy.
(unit: MeV).}
\begin{ruledtabular}
\begin{tabular}{lccc}
~~Channel   & Index & $\chi_J^{\sigma_i}$;~$\chi_I^{f_j}$;~$\chi_k^c$ & $M$~~ \\
        &   &$[i; ~j; ~k]$ &  \\[2ex]
$(B^* K^*)^1 (6217)$  & 1   & [1;~1;~3]  & $6226$  \\[2ex]
$(B^* K^*)^8$  & 2   & [1;~1;~3]   & $6449$   \\[2ex]
$(sb)^*(\bar{q}\bar{q})^*$  & 3   & [1;~1;~3]   & $6450$  \\[2ex]
$K_1$  & 4   & [1;~1;~3]   & $6437$  \\
            & 5   & [1;~1;~3]   & $6439$  \\[2ex]
$K_2$  & 6   & [1;~1;~3]   & $6298$  \\
             & 7   & [1;~1;~3]   & $6458$  \\[2ex]
$K_3$  & 8   & [1;~1;~3]   & $7093$  \\
             & 9   & [1;~1;~3]   & $6454$  \\[2ex]
$K_4$  & 10   & [1;~1;~3]   & $6441$  \\[2ex]
\multicolumn{3}{c}{Complete coupled-channels:} & $6226$
\end{tabular}
\end{ruledtabular}
\end{table}

\begin{figure}[ht]
\includegraphics[width=0.49\textwidth, trim={2.3cm 2.0cm 2.0cm 1.0cm}]{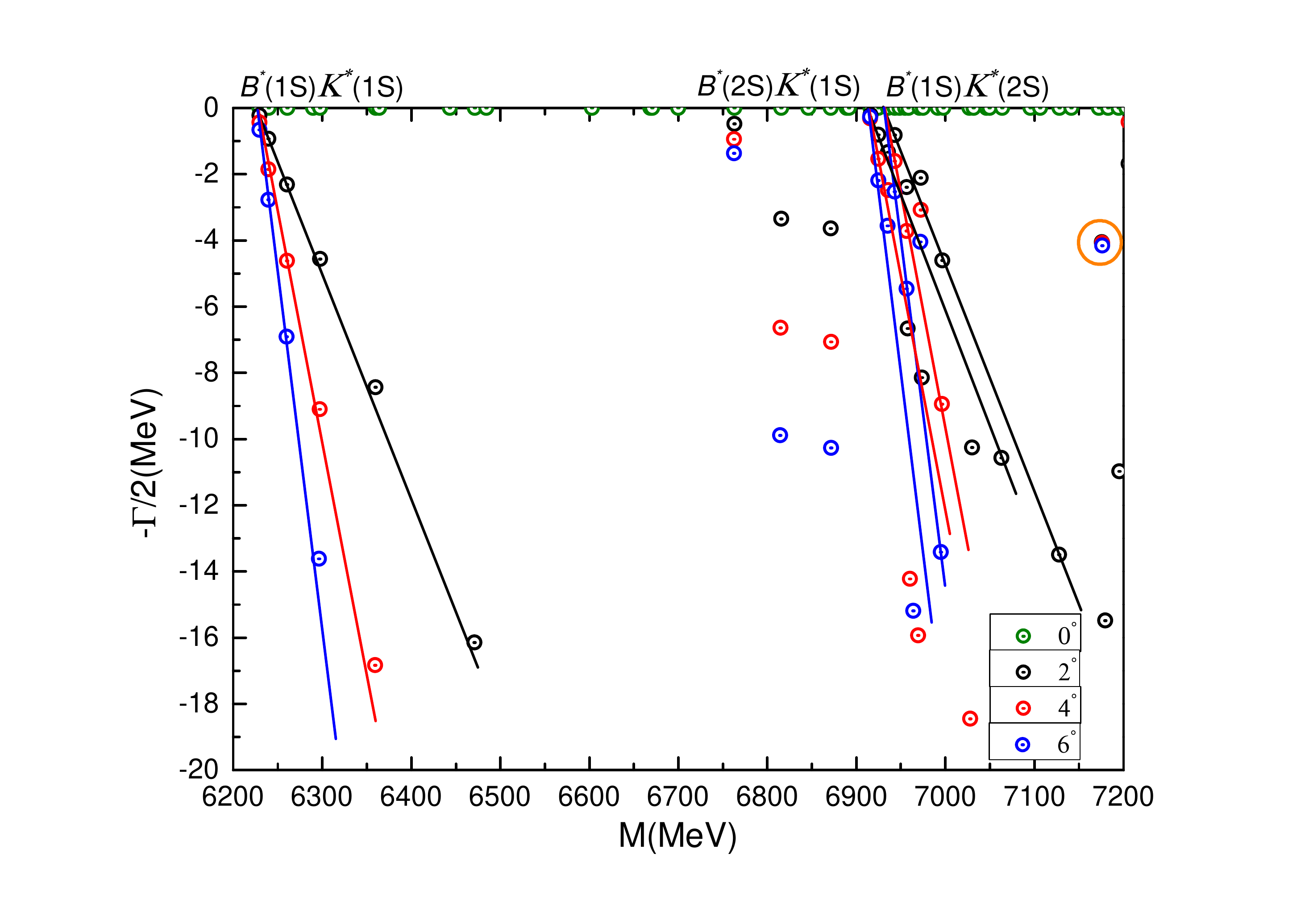}
\caption{\label{PP12} The complete coupled-channels calculation of the $sb\bar{q}\bar{q}$ tetraquark system with $I(J^P)=1(2^+)$ quantum numbers. We use the complex-scaling method of the chiral quark model varying $\theta$ from $0^\circ$ to $6^\circ$.}
\end{figure}

{\bf The $\bm{I(J^P)=1(2^+)}$ sector:} Table~\ref{GresultCC12} shows all the channels that can contribute to the mass of the highest spin and isospin state of $sb\bar{q}\bar{q}$ tetraquark. Again, no bound state is found in the single- and each-sector-coupled-channels calculations, with the lowest mass indicating the value of $B^*K^*$ threshold, 6226~MeV. Our results of the complete coupled-channels calculation in complex-range are shown in Fig.~\ref{PP12}. One can see that a $B^*(1S)K^*(2S)$ resonance state with mass 7175~MeV and width 8.1~MeV is obtained.


\begin{table}[!t]
\caption{\label{GresultCCT} Summary of the bound states and resonance structures found in the $sQ\bar{q}\bar{q}$ $(q=u,d,\,Q=c, b)$ tetraquark sectors. The first column shows the isospin, total spin and parity quantum numbers of each singularity, the second column refers to the dominant configuration, the obtained poles are presented with with the following notation: $E=M+i\Gamma$ in the last column, binding energy, $E_B$, is given in parentheses for the bound-states. (unit: MeV).}
\begin{ruledtabular}
\begin{tabular}{llc}
~ $I(J^P)$ & Dominant Channel   & Pole~~ \\
\hline
~~$0(0^+)$  & $sc\bar{q}\bar{q}$   & $2342\,(E_B=-36)$~~  \\
            & $sb\bar{q}\bar{q}$   & $5710\,(E_B=-49)$~~ \\
            & $B(1S)K(2S)$   & $6850+i2.0$~~ \\[2ex]
~~$0(1^+)$  & $sc\bar{q}\bar{q}$   & $2467\,(E_B=-31)$~~ \\
            & $sb\bar{q}\bar{q}$   & $5754\,(E_B=-46)$~~ \\
            & $B^*(1S)K^*(1S)$   & $6313+i12.5$~~ \\
            & $B^*(2S)K(1S)$   & $6754+i1.2$~~ \\[2ex]
~~$0(2^+)$  & $D^*(1S)K^*(1S)$   & $3493+i2.6$~~ \\
                  & $B^*(1S)K^*(1S)$   & $6754+i1.2$~~ \\[2ex]
~~$1(0^+)$  & $D^*(1S)K^*(1S)$   & $3089+i0.3$~~  \\
                  & $B(1S)K(2S)$   & $6876+i4.2$~~ \\[2ex]
~~$1(1^+)$  & $D^*(1S)K^*(1S)$   & $3091+i0.5$~~\\[2ex]
~~$1(2^+)$  & $D^*(1S)K^*(1S)$   & $3511+i2.1$~~ \\
                  & $B^*(1S)K^*(2S)$   & $7175+i8.1$~~
\end{tabular}
\end{ruledtabular}
\end{table}

\section{Summary}
\label{sec:summary}

The $sQ\bar{q}\bar{q}$ $(q=u,\,d,\, Q=c,\,b)$ tetraquarks with spin-parity $J^P=0^+$, $1^+$ and $2^+$, and in the iso-scalar and -vector sectors, are systemically investigated by means of real- and complex-scaling range of chiral quark model, along with a high efficiency numerical approach, Gaussian expansion method. The model, which contains one-gluon exchange, linear-screened confining and Goldstone-boson exchanges between light quarks interactions, has been successfully applied to the description of hadron, hadron-hadron and multiquark phenomenology. We considered in our calculations all possible tetraquark arrangements allowed by quantum numbers: singlet- and hidden-color meson-meson configurations, diquark-antidiquark arrangements with their allowed color triplet-antitriplet and sextet-antisextet wave functions, and four K-types structures.

Several bound states and resonance structures are found in both charm-strange $cs\bar{q}\bar{q}$ and bottom-strange $bs\bar{q}\bar{q}$ tetraquark sectors. Table~\ref{GresultCCT} summarizes our theoretical findings, we collect quantum numbers, dominant configuration and pole position in the complex plane for each singularity. We provide below a brief review of them.

Tightly bound charm- and bottom-strange tetraquarks are found in $I(J^P)=0(0^+)$ and $0(1^+)$ sectors. Their binding energies are around $-40$~MeV and they are interpreted as compact structures with all interquark distances $\lesssim1$~fm. Resonances of charm-strange tetraquarks are located within the energy interval 3.1-3.5 GeV, their total decay widths are usually less than 3 MeV and thus they are extremely narrow. Resonances of bottom-strange tetraquarks are located in the energy region 6.3-7.1 GeV with decay widths smaller than 13 MeV.

The development of this work was motivated by the recent LHCb's observation of two charmed-strange structures whose properties point out to be $ud\bar c\bar s$ tetraquarks. They were named $X_{0,1}(2900)$ and, within our formalism, are unstable with several candidates in the single-channel computations. We found 4 cases with $I(J^P)=0(0^+)$ quantum numbers: $K_1$ with mass at 2.89~GeV, $K_2$ at 2.96~GeV, $K_3$ at 2.97~GeV, and the hidden-color coupled-channels calculation delivering a mass 2.86~GeV. With quantum numbers $I(J^P)=0(1^+)$, only one candidate can be found in the hidden-color coupled-channels calculation with mass $2.94\,\text{GeV}$.


\begin{acknowledgments}
Work partially financed by: National Natural Science Foundation of China under Grant Nos. 11535005 and 11775118; the Ministerio Espa\~nol de Ciencia e Innovaci\'on under grant No. FIS2017-84114-C2-2-P; the Junta de Andaluc\'ia under contract No. Operativo FEDER Andaluc\'ia 2014-2020 UHU-1264517; but also PAIDI FQM-370.
\end{acknowledgments}


\bibliography{qsqQtetraquarks}

\end{document}